\documentclass[12pt, a4paper]{article} %scrartcl

\usepackage{setspace}
\usepackage[small, font = onehalfspacing]{caption}

\usepackage[authoryear]{natbib}
\bibpunct{(}{)}{;}{a}{,}{,}
\usepackage[english]{babel}
\usepackage[latin1]{inputenc}
\usepackage[T1]{fontenc}
\usepackage{lmodern}
\usepackage{amssymb}
\usepackage{xcolor,graphicx,xspace,caption}
\usepackage{listings,eso-pic,subfigure}
\usepackage{lscape}
\usepackage{multirow}
\usepackage{enumerate}
\usepackage{amsmath}
\pdfpageattr {/Group << /S /Transparency /I true /CS /DeviceRGB>>}

\definecolor{hint}{RGB}{191,63,0}

\definecolor{hellgelb}{rgb}{1,1,0.8}
\definecolor{colKeys}{rgb}{0,0,1}
\definecolor{colIdentifier}{rgb}{0,0,0}
\definecolor{colComments}{rgb}{1,0,0}
\definecolor{colString}{rgb}{0,0.5,0}

\usepackage{hyperref}
\hypersetup{linkcolor={blue}}

\newcommand{\E}{\mathop{\mbox{\sf E}}}
\newcommand{\PP}{\mathrm{P}}
\newcommand{\F}{\mathcal{F}}

\newcommand{\IF}{\mathbf{I}}

\title{lCARE - localizing Conditional AutoRegressive Expectiles
\footnote{Financial support from the Deutsche Forschungsgemeinschaft via CRC 649 ''Economic Risk'' and IRTG 1792 ''High Dimensional Non Stationary Time Series'', Humboldt-Universit\"{a}t zu Berlin, is gratefully acknowledged.}
\footnote{\textit{This is a post-peer-review, pre-copyedit version of an article published in the Journal of Empirical Finance. The final authenticated version is available online at:} http://dx.doi.org/10.1016/j.jempfin.2018.06.006}}
 \author{Xiu Xu\footnote{Humboldt-Universit\"{a}t zu Berlin,
C.A.S.E. - Center for Applied Statistics and Economics, Spandauer Str. 1, 10178 Berlin, Germany, tel: +49 (0)30 2093 5721, fax: +49 (0)30 2093 5649, Xiamen University, Wang Yanan Institute for Studies in Economics (WISE), 361005 Xiamen, China. Email: xiu.xu@hu-berlin.de}, Andrija Mihoci\footnote{Brandenburg University of Technology,
Chair of Economic Statistics and Econometrics, Erich Weinert Str. 1, 03046 Cottbus, Germany, tel: +49 (0)355 69 38 20}, Wolfgang Karl H\"{a}rdle\footnote{Humboldt-Universit\"{a}t zu Berlin,
C.A.S.E. - Center for Applied Statistics and Economics, Spandauer Str. 1, 10178 Berlin, Germany and School of Business, Singapore Management University, 50 Stamford Road, Singapore 178899}}
\date{\;}

\usepackage[top = 1in, bottom = 1in, left = 1in, right = 1in]{geometry}

\begin{document}

\maketitle
\vspace{-1.5cm}
\begin{abstract}
\footnotesize{\noindent

We account for time-varying parameters in the conditional expectile-based
value at risk (EVaR) model. The EVaR downside risk is more sensitive to the magnitude of portfolio losses compared to the quantile-based value at risk (QVaR). Rather than fitting the expectile models over ad-hoc fixed data windows, this study focuses on parameter instability of tail risk dynamics by utilising a local parametric approach. Our framework yields a data-driven optimal interval length at each time point by a sequential test. Empirical evidence at three stock markets from 2005-2016 shows that the selected lengths account for approximately
3-6 months of daily observations. This method performs favorable compared to the models with one-year fixed intervals, as well as quantile based candidates while employing a time invariant
portfolio protection (TIPP) strategy for the DAX, FTSE 100 and S\&P 500 portfolios. The tail risk
measure implied by our model finally provides valuable insights for asset allocation and portfolio
insurance.

$\;$

}

\end{abstract}

\newpage

\section{Introduction}
\label{Introduction}
Value at risk (VaR) is commonly used to measure the downside risk in finance, especially in portfolio risk management. Given a predetermined probability level, VaR evaluates the potential maximum loss for the targeted portfolio value; statistically it represents the quantile of the portfolio loss distribution, see \citet{Jorion:00}. Although it is straightforward to understand the VaR concept, it has been recently criticized. VaR lacks the property of sub-additivity, that is, under the VaR risk measure, the risk of a diversified portfolio may be larger than the sum of each individual asset risk, which in turn contradicts the common wisdom of diversification. \citet{Artzner:99} thus proposed the expected shortfall (ES) as a portfolio risk measure, i.e., the expected loss below a given threshold (e.g., VaR) given the risk probability level.

Another undesirable aspect of the VaR measure is its insensitivity to the magnitude of the portfolio loss. \citet{Kuan:09} provide an example where, under a given probability level, the potential downside risk changes under different tail loss distributions while the corresponding VaR remains the same. Since VaR merely depends on the probability value and neglects the size of the downside loss, \citet{Kuan:09} proposed a downside risk measure, the expectile-based Value at Risk (EVaR), a more sensitive measure of the magnitude of extreme losses than the conventional quantile-based VaR (QVaR). The expectile at given level is estimated by minimizing the asymmetric weighted least squared errors, exploring the method proposed by \citet{Newey:87}. The expectile level represents the relative cost of the expected margin shortfall, explained as the level of prudentiality. EVaR may be interpreted as a flexible QVaR \citep{Kuan:09}, because of the one-to-one mapping between quantiles and expectiles for a given loss distribution, see \citet{Efron:91}, \citet{Jones:94} and \citet{Yao:96}.

Models based on the expectile risk measure framework have thus been proposed, see e.g. \citet{Taylor:08} and \citet{Kuan:09} after \citet{Engle:04} successfully initialize the conditional autoregressive framework to model VaR. \citet{Kuan:09} moreover extend the EVaR to conditional EVaR and propose various Conditional AutoRegressive Expectile (CARE) specifications to accommodate stationary and weakly dependent data, extending the work by \citet{Newey:87}. Potential time-varying parameters resulting from the dynamic state of the economic and financial environment are however barely analysed. This is where this research comes into play. We focus on incorporating and reacting to potential structural breaks when estimating the expectile tail risk measure.

The proposed local parametric approach (LPA) utilizes a parametric model over adaptively chosen intervals. The essential idea of the LPA is to find the longest interval length guaranteeing a relatively small modelling bias, see e.g. \citet{Spokoiny:98} and \citet{Spokoiny:09}. The main advantage of the approach is the achievement of a balance between the modelling bias and parameter variability in data modelling. This approach has been successfully applied in many research areas: \citet{Cizek:09} analyse the GARCH$\left(1,1\right)$ models, \citet{Chen:10} explore it to forecast realised volatilities, \citet{Chen:14} predict the interest rate term structure, whereas \citet{Hardle:14a} utilise it successfully in high-frequency time series modelling and forecasting.

In this paper, we locally estimate the expectile risk measure rather than following a traditional approach of assuming constant CARE parameters over (relatively long) ad-hoc selected data intervals. Based on one of the conditional expectile model specifications in \citet{Kuan:09} and assuming that the error term follows the asymmetric normal distribution, \citet{Gerlach:12} and \citet{Gerlach:14}, we dynamically estimate the time-varying CARE parameters over potentially varying intervals of homogeneity. The desired intervals of homogeneity are found by a sequential testing procedure. The resulting (time-varying) interval lengths indicate the presence of potential structural changes in tail risk measurement.

It is worth mentioning that several articles consider the dynamic window selection of time-varying parameters, \citet{Pesaran:07} and \citet{Inoue:14}, or introduce varying-coefficient models for tail risk measure estimation, \citet{Honda:04}, \citet{Kim:07} and \citet{Cai:08}. Most of the research, however, mainly explores nonparametric approaches or considers polynomial splines to estimate the conditional quantile. A state space signal extraction algorithm has been applied to compute spline-based non-parametric quantile and expectile regressions by \citet{De:09}, while \citet{Xie:14} develop a nonparametric varying-coefficient approach to model the expectile-based value at risk.

In our research it turns out that the proposed localised conditional autoregressive expectile (lCARE) model successfully captures tail risk dynamics by taking the time-varying parameter characteristics and potential market condition structure changes into account while measuring the risk associated with tail events. Based on empirical results, we find that at the $0.25\%$  expectile level the typical interval lengths that strike a balance between bias and variability in daily time series include on average 100 days. At the higher, $5\%$ expectile level, the selected interval lengths range roughly between 80-90 days. The resulting time-varying expectile series allows us moreover to consider the dynamics of other tail risk measures, most prominently that of quantiles or the expected shortfall.

The methodology presented here is successfully applied to a portfolio insurance strategy for the DAX, FTSE 100 and S\&P 500 index portfolios. A portfolio insurance strategy is designed to guarantee a minimum asset portfolio value over a selected investment horizon, where the downside risk can be reduced and controlled while investors can participate in the potential gains. The proportion of the value invested into the risky asset (here the selected index portfolio), denoted as the multiplier, is directly related to the estimated tail risk measure.
A standard approach keeps the multiplier fixed regardless of the market conditions, \citet{Estep:88}, \citet{Hamidi:14}, whereas we exercise the protection strategy utilising the dynamic tail risk measure implied by the lCARE model. Comparison to the benchmarks - one-year fixed rolling window CARE estimation and quantile-based (CAViaR) estimation - reveals that the lCARE model presents a striking
outperformance in portfolio insurance.

This paper is structured as follows: firstly, the data is presented in Section \ref{Data} whereas Section \ref{LCARE} introduces the lCARE modelling framework based on the CARE setup and the local parametric approach in tail risk modelling. Section \ref{EMP} presents the empirical results and finally, Section \ref{Conclusions} concludes.

\section{Data}
\label{Data}

\begin{figure}
\begin{center}
\includegraphics[scale = 0.6]{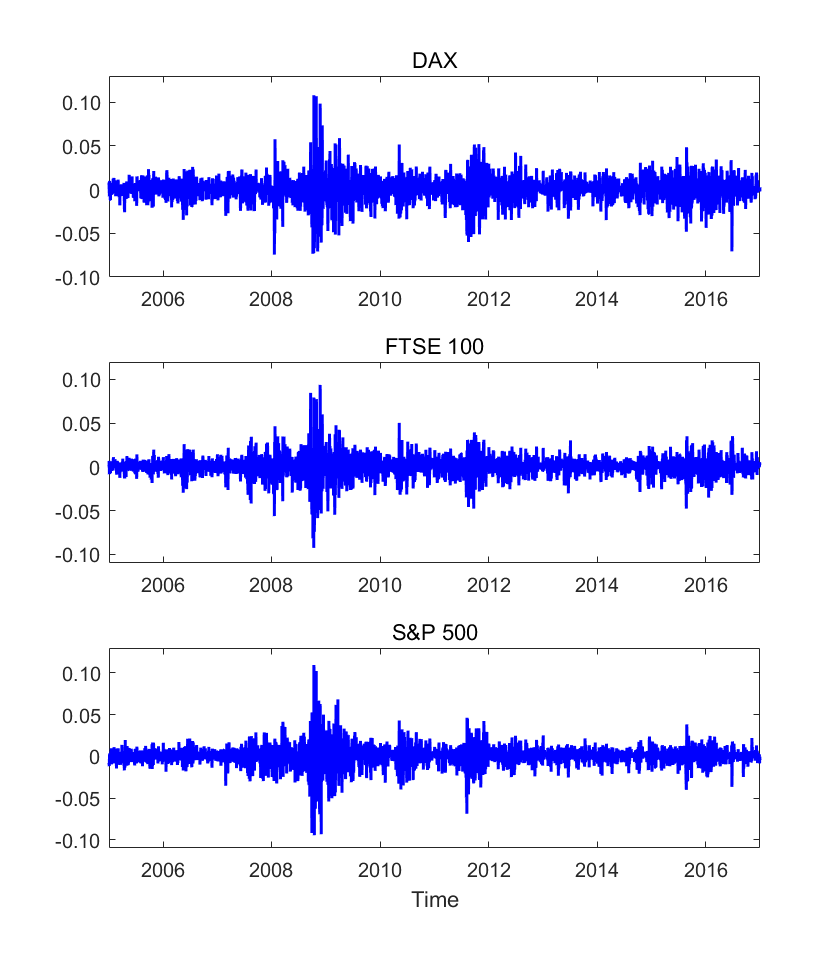}
\caption{Selected index return time series from 3 January 2005 to 30 December 2016 (3130 trading days). %\hspace*{\fill}\raisebox{-1pt}{\includegraphics[scale=0.05]{qletlogo}}
%\href{https://github.com/QuantLet/lCARE-BTU-HUB/tree/master/LCARE_Index_Returns}{LCARE\_Index\_Returns}
}
\label{CARE_return_figure}
\end{center}
\end{figure}

In risk modelling we consider three stock markets and focus on the dynamics of the representative index time series, namely, the DAX, FTSE 100 and S\&P 500 series. Daily index returns are obtained from Datastream and our data cover the period from 3 January 2005 to 30 December 2016, in total 3130 trading days. The daily returns evolve similarly across the selected markets and all present relatively large variations during the financial crisis period from 2008-2010, see Figure \ref{CARE_return_figure}. Although the return time series exhibit nearly zero-mean with slightly pronounced skewness values, all present comparatively high kurtosis, see Table \ref{Return_summary} that collects the summary statistics. Please note that the empirical results of this paper as well as the corresponding MATLAB programming codes can be found in the folder \href{https://github.com/QuantLet/lCARE-BTU-HUB}{https://github.com/QuantLet/lCARE-BTU-HUB} as well as at \href{http://quantlet.de/d3/ia/}{http://quantlet.de/d3/ia/}.

\vspace{-0.20cm}

\begin{table}[htp]
\begin{center}
\begin{tabular}{l|r@{.}lr@{.}lr@{.}lr@{.}lr@{.}lr@{.}lr@{.}l}
\hline\hline
\multicolumn{1}{c}{Index} \vline	&	\multicolumn{2}{c}{\makebox[1.2cm][c]{Mean}}	&	\multicolumn{2}{c}{\makebox[1.2cm][c]{Median}}	&	\multicolumn{2}{c}{\makebox[1.2cm][c]{Min}}	&	\multicolumn{2}{c}{\makebox[1.2cm][c]{Max}}	&	\multicolumn{2}{c}{\makebox[1.2cm][c]{Std}}	&	\multicolumn{2}{c}{\makebox[1.2cm][c]{Skew.}}	&	\multicolumn{2}{c}{\makebox[1.2cm][c]{Kurt.}}	\\
\hline
%DAX	&	0 & 0003 	&	0 & 0007 	 	&	-0 & 0743 &	0 & 1080	&	0 & 0137 	&	0 & 0357 	&	10 & 1654 	\\
%FTSE 100	&	0 & 0001 	&	0 & 0001 	 	&	-0 & 0927 &	0 & 0938	&	0 & 0120 	&	-0 & 1498 	&	11 & 9066 	\\
%S\&P 500	&	0 & 0002 	&	0 & 0005 	 	&	-0 & 0947 &	0 & 1096	&	0 & 0127 	&	-0 & 3364 	&	14 & 5131    \\

DAX  &  0&0003  &  0&0007  &  0&1080  & -0&0743  &  0&0137  & -0&0406  &  9&2297  \\
FTSE 100 &  0&0001  &  0&0001  &  0&0938  & -0&0927  &  0&0117  & -0&1481  & 11&2060  \\
S\&P 500 &  0&0002  &  0&0003  &  0&1096  & -0&0947  &  0&0121  & -0&3403  & 14&6949  \\
\hline\hline
\end{tabular}
\caption{Descriptive statistics for the selected index return time series from 3 January 2005 to 30 December 2016 (3130 trading days): mean, median, minimum (Min), maximum (Max), standard deviation (Std), skewness (Skew.) and kurtosis (Kurt.).
%\hspace*{\fill} %\raisebox{-1pt}{\includegraphics[scale=0.05]{qletlogo}}\href{https://github.com/QuantLet/lCARE-BTU-HUB/tree/master/LCARE_Index_Returns_Description}{LCARE\_Index\_Returns\_Description}
}
\label{Return_summary}
\end{center}
\end{table}

\section{Localized Conditional Autoregressive Expectiles}
\label{LCARE}

Understanding tail risk plays an essential role in asset pricing, portfolio allocation, investment performance evaluation and external regulation. Tail event dynamics is commonly assessed through the employment of parametric, semi-parametric or nonparametric techniques, see, e.g., \citet{Taylor:08}. Our paper contributes to the econometric literature by localizing parametric CARE specifications by \citet{Kuan:09} and explores the effects of potential market structure changes when modelling tail risk measures. In this section we summarise the current research on expectile-based risk management and conduct a detailed empirical study of the parameter dynamics of the introduced DAX, FTSE 100 and S\&P 500 return series. The results motivate the application of the local parametric approach by \citet{Spokoiny:98} and finally the localized conditional autoregressive expectile (lCARE) model provides a sound downside risk assessment framework for quantitative finance practice.

\subsection{Conditional Autoregressive Expectile Model}
\label{LCARE1}

Tail risk exposure can successfully be captured by an expectile-based risk measure, in contrast to modelling risk solely using Value at Risk (VaR). Despite being the most commonly used (not coherent) tail risk measure, VaR exhibits insensitivity to the potential magnitude of the loss, see, e.g., \citet{Acerbi:02}, \citet{Taylor:08}. After the conditional autoregressive value at risk (CAViaR) model  by \citet{Engle:04} was proposed, \citet{Taylor:08} found that VaR, based on the conditional autoregressive expectile model, is more sensitive to the underlying tail risk distribution. The conditional autoregressive expectile (CARE) model specifications by \citet{Kuan:09} nevertheless directly model the return time series and extend the asymmetric least square estimation method by \citet{Newey:87} in the analysis of stationary but weakly dependent time series data.

CARE model specifications provide insights into the dynamics of financial data and offer valuable economic interpretation. Although quantiles and expectiles belong to M-quantiles, see, e.g., \citet{Jones:94}, the implications in risk assessment differ considerably. VaR is a zero-moment whereas expectile is a first-moment tail risk measure, thus in the former case the proportion of asymmetric downside and upside quantile level is determined only by the ratio between downside and upside probabilities. Expectiles measure the proportion of asymmetric downside and upside expectile level while capturing the ratio between the expected marginal shortfall. Equivalently, the potential cost of more extreme losses and the opportunity cost due to the expected marginal overcharge is captured by expectiles. The CARE specifications furthermore accommodate stylised facts of the return time series, such as weak serial dependence, or volatility heteroskedasticity. Accommodating asymmetric effects on the tail expectiles of the positive and negative returns becomes essential in interpreting tail risk dynamics.

Consider the CARE model specification for a return time series $y = \left\{y_{t}\right\}^{n}_{t = 1}$
%\vspace{-0.20cm}
\begin{align}
\begin{split}
 y_{t}  = & \, e_{t, \tau} + \varepsilon_{t, \tau} \\
 e_{t, \tau}  = & \, \alpha_{0, \tau} + \alpha_{1, \tau}y_{t - 1} + \alpha_{2, \tau}\left(y^{+}_{t - 1}\right)^{2} + \alpha_{3, \tau}\left(y^{+}_{t - 2}\right)^{2} + \alpha_{4, \tau}\left(y^{+}_{t - 3}\right)^{2} + \\
 & \, + \alpha_{5, \tau}\left(y^{-}_{t - 1}\right)^{2}+ \alpha_{6, \tau}\left(y^{-}_{t - 2}\right)^{2}+ \alpha_{7, \tau}\left(y^{-}_{t - 3}\right)^{2}
\label{LCARE_Expectile_Specification}
\end{split}
\end{align}
where $e_{t, \tau}$ and $\varepsilon_{t, \tau}$ denote the expectile and the error term at level $\tau \in \left(0, 1\right)$ and time $t$, respectively. For $j = 1, 2, 3$, $y^{+}_{t - j} = \max\left\{y_{t - j}, 0\right\}$ and $y^{-}_{t - j} = \min\left\{y_{t - j}, 0\right\}$ denote the positive or negative observed $j$-th period lagged returns at time $t$, respectively.

The $\tau$-level expectile $e_{t, \tau}$ in Equation \eqref{LCARE_Expectile_Specification} can be estimated by minimising the asymmetric least square (ALS) loss function
\begin{equation}
\sum_{t = 4}^{n}\left|\tau - \IF\left(y_{t} \leq e_{t, \tau}\right)\right|\left(y_{t} - e_{t, \tau}\right)^{2}
\label{CARE_ALS}
\end{equation}
with $\IF\left(\cdot\right)$ denoting the indicator function.

Within the CARE framework, \citet{Gerlach:14} and \citet{Gerlach:12} assume that the error term $\varepsilon_{t, \tau}$ follows the asymmetric normal distribution (AND). We assume that, conditional on the information set $\F_{t - 1}$, the data process follows an asymmetric normal distribution AND$\left(\mu, \sigma^{2}_{\varepsilon_{\tau}}, \tau\right)$ with pdf:
\begin{equation}
f\left(y_{t} - \mu \left.\right|\F_{t - 1}\right) = \displaystyle\frac{2}{\sigma_{\varepsilon_{\tau}}} \left(\sqrt{\displaystyle\frac{\pi}{\left|\tau - 1\right|}} + \sqrt{\displaystyle\frac{\pi}{\tau}}\right)^{-1} \exp\left\{-\eta_{\tau}\left(\displaystyle\frac{y_{t} - \mu}{\sigma_{\varepsilon_{\tau}}}\right)\right\}
\label{lCARE_fvarepsilon}
\end{equation}
where $\eta_{\tau}\left(u\right) = \left|\tau - \IF\left\{u \leq 0\right\}\right|u^{2}$ is the employed check function, $\mu$ represents the expectile value to be estimated and $\sigma^{2}_{\varepsilon_{\tau}}$ denotes the variance of the error term. It is worth noting that maximising the likelihood based on the distribution \eqref{lCARE_fvarepsilon} is mathematically equivalent to minimising the asymmetric least square loss function \eqref{CARE_ALS}.

Conditional on the information set $\F_{t - 1}$ up to observation $\left(t - 1\right)$, the expectile $e_{t, \tau}$ includes a lagged return component and it mimics several financial series features, namely, the volatility clustering and potential asymmetric magnitude effects. Note that at level $\tau = 0.5$, the expectile equals to the mean value. Given specification \eqref{CARE_ALS}, the parameter vector finally contains nine elements, namely\\
\vspace{-1.00cm}
%\begin{center}
%$\theta_{\tau} = \left(\alpha_{0, \tau}, \alpha_{1, \tau}, \alpha_{2, \tau}, \alpha_{3, \tau}, \alpha_{4, \tau}, \alpha_{5, \tau}, \alpha_{6, \tau}, \alpha_{7, \tau}, \sigma^{2}_{\varepsilon_{\tau}}\right)^{\top}$.
%\end{center}

\begin{center}
$\theta_{\tau} = \left(\alpha_{0, \tau}, \alpha_{1, \tau}, \alpha_{2, \tau}, \alpha_{3, \tau}, \alpha_{4, \tau}, \alpha_{5, \tau}, \alpha_{6, \tau}, \alpha_{7, \tau}, \sigma_{\varepsilon_{\tau}}\right)^{\top}$.
\end{center}

In specification \eqref{LCARE_Expectile_Specification}, the parameter $\alpha_{1, \tau}$ indirectly measures the persistence level in the conditional expectile tail through the lagged return series. Since the parameters related to the past positive or negative squared returns potentially differ, specification \eqref{LCARE_Expectile_Specification} accounts for the asymmetric effects of the positive and negative squared lagged returns on the conditional tail expectile magnitude. This similarly mimics the leverage effect associated with volatility modelling, where negative (positive) returns are followed by relatively larger (lower) variability. Under the working assumption that the expectile tail dynamics can be well approximated over a given data interval by a model with constant parameters, it suffices to include three lags in modelling return series.

The resulting quasi log likelihood function for observed data $\mathcal{Y} = \left\{y_{1}, \ldots, y_{n}\right\}$ over a fixed interval $I$ is given by
\vspace{0.10cm}
\begin{equation}
\ell_{I}\left(\mathcal{Y}; \theta_{\tau}\right) = \displaystyle\sum_{t \in I} \log f\left(y_{t} - e_{t, \tau}\left.\right|\F_{t - 1}\right)
\label{CARE_likelihood}
\vspace{0.10cm}
\end{equation}
The quasi maximum likelihood estimate (QMLE) for the CARE parameter is then obtained through
\begin{equation}
\widetilde{\theta}_{I, \tau} = \operatorname{arg}\,\underset{\theta_{\tau} \in \Theta}{\operatorname{max}} \, \ell_{I}\left(\mathcal{Y}; \theta_{\tau}\right)
\label{LCARE_QMLE}
\end{equation}
over a right-end fixed interval $I = \left[ t_{0} - m, t_{0} \right]$ of $\left(m + 1\right)$ observations at observation $t_{0}$.

\subsection{Parameter Dynamics}
\label{PD}

The idea behind the local parametric approach (LPA) is to find the optimal (in-sample) data interval over which one can safely fit a parametric model with time-invariant parameters. This optimal interval, the so-called interval of homogeneity, is selected among pre-specified right-end interval candidates at each time point. The proposed lCARE model is thus able to incorporate potential structure breaks in expectile dynamics. In this part we implement a fixed rolling window exercise in order to provide empirical evidence on the time-varying characteristics of the CARE estimates, as well as to select the 'true' parameter constellation used in the LPA simulation. At the end we discuss the estimation quality of the QMLE \eqref{LCARE_QMLE}.

\textbf{Dynamics and Distributional Characteristics}

In the analysis of the selected (daily) stock market indices presented in Section \ref{Data}, we consider different interval lengths (e.g., 60, 125 and 250 observations) and analyse the corresponding estimates. One may observe a relatively large variability of the estimated parameters while fitting the model over short data intervals and vice versa. Note that the modelling bias moves in the opposite direction: shorter (longer) intervals lead to a relatively low (high) modelling bias. The distributional features of the estimated CARE parameters are here studied through three expectile level cases, namely $\tau = 0.0025$, $\tau = 0.01$ and $\tau = 0.05$. Our conducted rolling window estimation exercise finally provides valuable insights into the expectile (distribution) dynamics.

\begin{figure}[ht]
\begin{center}
\includegraphics[scale = 0.60]{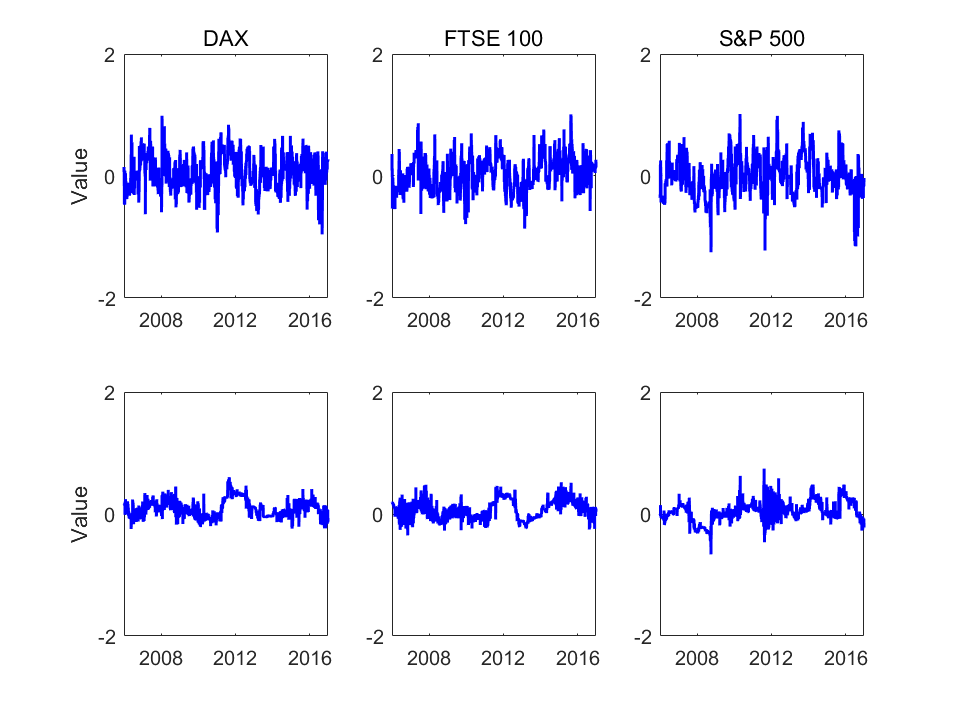}
\caption{Estimated parameter $\widetilde{\alpha}_{1, 0.05}$ across the three selected stock markets from 2 January 2006 to 30 December 2016, with 60 (upper panel) and 250 (lower panel) observations used in the rolling window exercise at fixed expectile level $\tau = 0.05$.
%\hspace*{\fill} \raisebox{-1pt}{\includegraphics[scale=0.05]{qletlogo}}\href{https://github.com/QuantLet/lCARE-BTU-HUB/tree/master/LCARE_Estimation_Rolling_005}{LCARE\_Estimation\_Rolling\_005}
}
\label{CARE_estimate_rolling_figure_005}
\end{center}
\end{figure}

\begin{figure}[ht]
\begin{center}
\includegraphics[scale = 0.60]{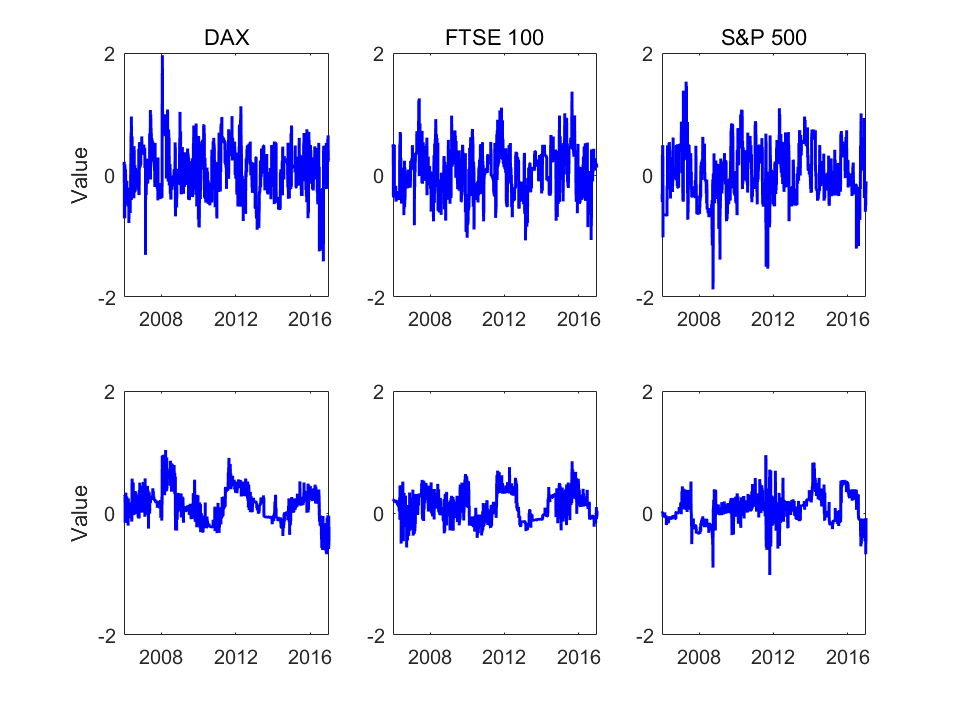}
\caption{Estimated parameter $\widetilde{\alpha}_{1, 0.01}$ across the three selected stock markets from 2 January 2006 to 30 December 2016, with 60 (upper panel) and 250 (lower panel) observations used in the rolling window exercise at fixed expectile level $\tau = 0.01$.
%\hspace*{\fill} \raisebox{-1pt}{\includegraphics[scale=0.05]{qletlogo}}\href{https://github.com/QuantLet/lCARE-BTU-HUB/tree/master/LCARE_Estimation_Rolling_001}{\,LCARE\_Estimation\_Rolling\_001}
}
\label{CARE_estimate_rolling_figure_001}
\end{center}
\end{figure}

\begin{figure}[ht]
\begin{center}
\includegraphics[scale = 0.60]{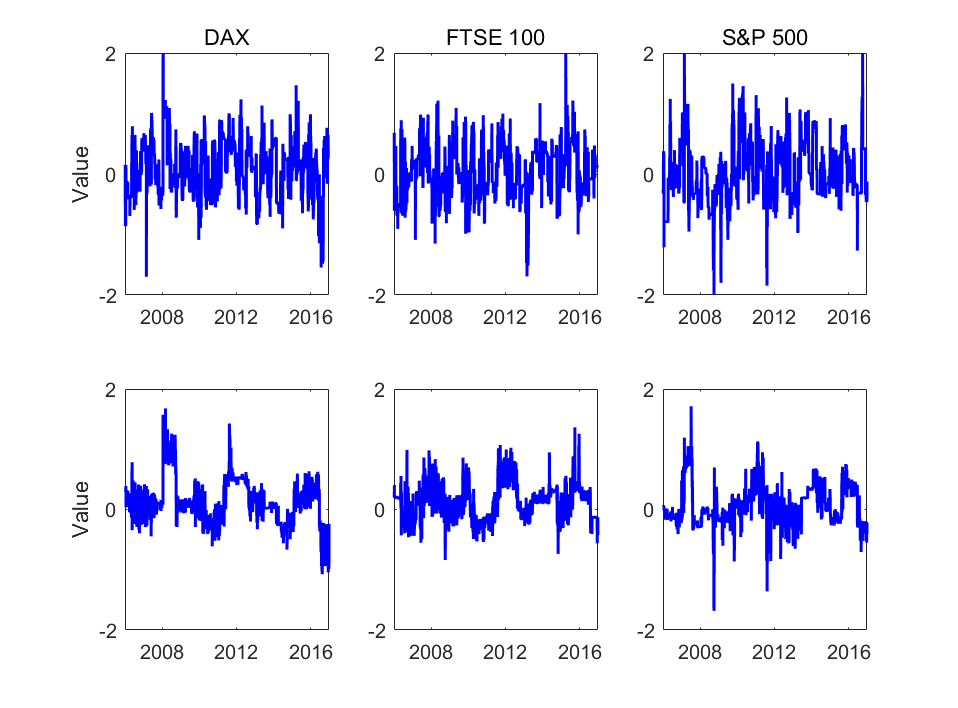}
\caption{Estimated parameter $\widetilde{\alpha}_{1, 0.0025}$ across the three selected stock markets from 2 January 2006 to 30 December 2016, with 60 (upper panel) and 250 (lower panel) observations used in the rolling window exercise at fixed expectile level $\tau = 0.0025$.
%\hspace*{\fill} \raisebox{-1pt}{\includegraphics[scale=0.05]{qletlogo}}\href{https://github.com/QuantLet/lCARE-BTU-HUB/tree/master/LCARE_Estimation_Rolling_005}{LCARE\_Estimation\_Rolling\_005}
}
\label{CARE_estimate_rolling_figure_00025}
\end{center}
\end{figure}

Parameter estimates are indeed more volatile while fitting the data over shorter intervals with a comparably smaller modelling bias as compared to schemes using longer window sizes, see e.g. Figures \ref{CARE_estimate_rolling_figure_005}, \ref{CARE_estimate_rolling_figure_001} and \ref{CARE_estimate_rolling_figure_00025}. Here we display the estimated CARE parameters $\widetilde{\alpha}_{1, 0.05}$, $\widetilde{\alpha}_{1, 0.01}$ and $\widetilde{\alpha}_{1, 0.0025}$ in a rolling window exercise across the three selected stock market indices from 2 January 2006 to 30 December 2016 at levels $\tau = 0.05$, $\tau = 0.01$ and $\tau = 0.0025$, respectively. The upper (lower) panel at each figure shows the estimated parameter values if 60 (250) observations are included in the respective window.

Supportive evidence for the variance-bias trade-off is furthermore provided by the density estimates of the parameters involved across all three analysed three stock market indices. Kernel density plots (using a Gaussian kernel with optimal bandwidth) of estimated parameters show that shorter intervals again lead to more variability of the estimates and vice versa. For the sake of brevity we refrain from showing the density estimates. The densities are quite distinct in the two extreme cases (60 vs 250 observations).

\textbf{Descriptive Statistics}

The lCARE testing framework demands a set of simulated critical values that rely on reasonable parameter constellations. A data driven approach to select the 'true' parameters is in our work based on a sample window covering one year, i.e., 250  observations. Descriptive statistics of the resulting estimated CARE parameters across all three investigated time series from 2 January 2006 to 30 December 2016 (2870 trading days) are provided in Table \ref{CARE_parameter_dynamics_quartiles}. The first quartile of all estimated parameter values is labelled as 'low', the median as 'mid' and the third quartile as 'high'. Note that the analysis has been conducted at three expectile levels, $\tau = 0.0025$, $\tau = 0.01$ and $\tau = 0.05$, and that at a given expectile level, there are three 'true' parameter constellations, i.e., parameter values most likely found in practice.

\begin{table}[htp]
\begin{center}
\begin{tabular}{c|r@{.}lr@{.}lr@{.}l|r@{.}lr@{.}lr@{.}l|r@{.}lr@{.}lr@{.}l}
\hline\hline
%\multirow{2}{*}{Model}
 & \multicolumn{6}{c}{$\tau = 0.05$}  \vline & \multicolumn{6}{c}{$\tau = 0.01$} \vline & \multicolumn{6}{c}{$\tau = 0.0025$}\\
 & \multicolumn{2}{c}{Low} & \multicolumn{2}{c}{Mid} & \multicolumn{2}{c}{High} \vline
 & \multicolumn{2}{c}{Low} & \multicolumn{2}{c}{Mid} & \multicolumn{2}{c}{High} \vline
 & \multicolumn{2}{c}{Low} & \multicolumn{2}{c}{Mid} & \multicolumn{2}{c}{High}\\
\hline

$\widetilde{\alpha}_{0, \tau}$ & -0&016   & -0&013   & -0&009   & -0&026   & -0&021   & -0&015   & -0&034   & -0&026   & -0&021   \\
$\widetilde{\alpha}_{1, \tau}$ & -0&035   &  0&051   &  0&153   & -0&075   &  0&079   &  0&240   & -0&131   &  0&090   &  0&295   \\
$\widetilde{\alpha}_{2, \tau}$ &  0&079   &  0&145   &  0&209   &  0&077   &  0&155   &  0&247   & -0&319   &  0&120   &  0&207   \\
$\widetilde{\alpha}_{3, \tau}$ &  0&037   &  0&138   &  0&247   & -0&074   &  0&162   &  0&232   & -0&116   &  0&152   &  0&561   \\
$\widetilde{\alpha}_{4, \tau}$ &  0&052   &  0&147   &  0&246   &  0&101   &  0&170   &  0&452   &  0&062   &  0&152   &  0&740   \\
$\widetilde{\alpha}_{5, \tau}$ &  0&004   &  0&115   &  0&244   & -0&055   &  0&106   &  0&308   &  0&031   &  0&141   &  1&463   \\
$\widetilde{\alpha}_{6, \tau}$ &  0&022   &  0&104   &  0&156   & -0&576   &  0&109   &  0&160   & -1&748   &  0&113   &  0&179   \\
$\widetilde{\alpha}_{7, \tau}$ & -0&014   &  0&099   &  0&152   & -0&861   &  0&106   &  0&149   & -3&124   &  0&108   &  0&161   \\
 $\widetilde{\sigma}_{\varepsilon_{\tau}}$ & 0 & 001	&	0 & 002	&	0 & 003	&	0 & 001	&	0 & 002	&	0 & 002	&	0 & 001	&	0 & 002	&	0 & 001 \\
\hline\hline
\end{tabular}
\caption{Descriptive statistics of estimated CARE parameters. All estimated CARE parameters based on the window covering one year, i.e., 250 observations, for the three stock market indices from 2 January 2006 to 30 December 2016 (2870 trading days) are pooled together for the two expectile levels $\tau = 0.05$, $\tau = 0.01$ and $\tau = 0.0025$, respectively. We label the first quartile as 'low', the median as 'mid' and the third quartile as 'high'.  %\hspace*{\fill}\raisebox{-1pt}{\includegraphics[scale=0.05]{qletlogo}}\href{https://github.com/QuantLet/lCARE-BTU-HUB/tree/master/LCARE_Parameter_Dynamics_Quartiles}
%{\,LCARE\_Parameter\_Dynamics\_Quartiles}
}
\label{CARE_parameter_dynamics_quartiles}
\end{center}
\end{table}

\textbf{Estimation Quality}

%The estimation quality of the quasi-maximum likelihood approach is addressed here. Denote the pseudo \textit{true} parameter vector as $\theta^{\ast}_{\tau}$ at expectile level $\tau$, the quality of estimating $\theta_{\tau}^{\ast}$ by quasi-maximum likelihood estimator (QMLE) $\widetilde{\theta}_{I, \tau}$ given in \eqref{LCARE_QMLE} is measured in terms of the Kullback-Leibler divergence
%\begin{equation}
%\operatorname{\E}_{\theta^{\ast}_{\tau}}\left|\ell_{I}\left(\mathcal{Y}; \widetilde{\theta}_{I, \tau}\right)-\ell_{I}\left(\mathcal{Y}; \theta^{\ast}_{\tau}\right)\right|^{r} \leq \mathcal{R}_{r}\left(\theta^{\ast}_{\tau}\right)
%\label{CARE_KL}
%\end{equation}
%with $\mathcal{R}_{r}\left(\theta^{\ast}_{\tau}\right)$ denoting the risk bound, see, e.g., \citet{Mercurio:04} and \citet{Spokoiny:09}. In practice the modest risk power $r = 0.5$ leads to relatively shorter intervals of homogeneity as compared with the conservative risk case with $r = 1$. According to the pseudo true parameter vector, we simulate thousand time series of the CARE specification and take the largest average value of the ($r$-th power) difference between the respective log-likelihood values, see equation \eqref{CARE_KL}, as the corresponding risk bound. Note that the considered interval candidates in this simulation covered
%\vspace{-0.50cm}
%\begin{center}
% $\left\{20, 25, 31, 39, 49, 61, 76, 95, 119, 149, 186, 250\right\}$
%\end{center}
%\vspace{-0.50cm}
%observations - see the selection details in the following sub-section.

Here we address the estimation quality of the quasi-maximum likelihood approach. Denote the pseudo-true parameter vector at expectile level $\tau$ as $\theta^{\ast}_{\tau}$. The quality of estimating the unknown parameter by the quasi-maximum likelihood estimator (QMLE) $\widetilde{\theta}_{I, \tau}$ given in \eqref{LCARE_QMLE} is measured in terms of the Kullback-Leibler divergence
\begin{equation}
\operatorname{\E}_{\theta^{\ast}_{\tau}}\left|\ell_{I}\left(\mathcal{Y}; \widetilde{\theta}_{I, \tau}\right)-\ell_{I}\left(\mathcal{Y}; \theta^{\ast}_{\tau}\right)\right|^{r} \leq \mathcal{R}_{r}\left(\theta^{\ast}_{\tau}\right)
%\operatorname{\E}_{\theta^{\ast}_{\tau}}\left|\ell_{I}\left(\mathcal{Y}; \widetilde{\theta}_{I, \tau}\right)-\ell_{I}\left(\mathcal{Y}; \theta^{\ast}_{\tau}\right)\right|^{r} \leq \mathcal{R}_{r}\left(\theta^{\ast}_{\tau}\right)
\label{CARE_KL}
\end{equation}
with $\mathcal{R}_{r}\left(\theta^{\ast}_{\tau}\right)$ denoting the risk bound, see, e.g., \citet{Mercurio:04} and \citet{Spokoiny:09}. In the selection of the power risk level $r$, we follow empirical evidence. In recent studies, a lower selected risk power level $r$ leads to relatively shorter intervals of homogeneity and vice versa, thus it is recommended to consider the moderate risk case ($r = 0.5$ or $r = 0.8$) or the so-called conservative risk case, $r = 1$, see \citet{Hardle:14a}. Since our results favour the conservative risk case (results for $r = 0.5$ are not reported here), in further work we use the risk powers $r = 0.8$ and $r = 1$.
The parametric risk bound is determined through a simulation study presented in Appendix \ref{Appendix_01}.
%The values of the simulated risk bound $\mathcal{R}_{r}\left(\theta^{\ast}_{\tau}\right)$ according to equation \eqref{CARE_KL} across different setups is given in Table \ref{CARE_risk_bound}. We consider the modest $\left(r = 0.50\right)$ and the conservative $\left(r = 1.00\right)$ risk case for two expectile levels,  $\tau = 0.05$ and $\tau = 0.01$. The risk bounds are obtained by Monte Carlo simulation for each selected parameter vector corresponding to Table \ref{CARE_parameter_dynamics_quartiles} where we label the first quartile of estimated parameters as 'low', the mean as 'mid' and the third quartile as 'high'. It turns out that the risk bounds in the conservative risk case ($r = 1$) are relatively larger than the bounds obtained in the modest risk case with $r = 0.5$.

%Key empirical results from the presented fixed rolling window exercise can be summarised as follows: (a) with different estimation sample windows, a tradeoff between the modelling bias and parameter variability exists, (b) the estimated parameter characteristics as well as the estimation quality results demand a method that successfully accommodates time-varying parameters, (c) data intervals covering 60 to 250 observations may be suitable in providing a good balance between the bias and variability, (d) it is reasonable practice to select three data-driven 'true' parameter constellations for each expectile level in daily risk management. Motivated by these findings, we now introduce some more details of lCARE.

Key empirical results from the presented fixed rolling window exercise can be summarised as follows: (a) there exists a trade-off between the modelling bias and parameter variability across different estimation setups, (b) the characteristics of the time series of estimated parameter values as well as the estimation quality results demand the application of an adaptive method that successfully accommodates time-varying parameters, (c) data intervals covering 60 to 250 observations may provide a good balance between the bias and variability, (d) it is reasonable to select three data-driven 'true' parameter constellations for each expectile level in daily risk management. Motivated by these findings, we now introduce our lCARE modelling framework.

\subsection{Localizing Conditional Autoregressive Expectile Model }
\label{lCARE}

How to account for the time-varying characteristics of CARE parameters in tail risk modelling? Here we utilize the aforementioned local parametric approach (LPA), which has been gradually introduced to time series literature. The essential idea of the proposed lCARE framework is to find the longest time series data interval over which the CARE model can be approximated by a specification with time-invariant parameters. This interval is labelled as the interval of homogeneity. By a sequential testing procedure, the so-called local change point detection test, we adaptively select the interval of homogeneity among interval candidates. The critical values of the sequential test are simulated by a Monte Carlo method, for details we refer to Appendix \ref{Appendix_02}. Finally, the adaptively estimated parameter vector at every time point (for example, at each trading day) is selected based on the test outcome.

\textbf{Interval Selection}

There are many possible candidates for these intervals of homogeneity. To alleviate the computational burden, we choose $\left(K + 1\right)$ nested intervals of length $n_{k} = \left|I_{k}\right|$, $k = 0, \ldots, K$, i.e., $I_{0} \subset I_{1} \subset \cdots \subset I_{K}$. Interval lengths are assumed to be geometrically increasing with $n_{k} = \left[n_{0}c^{k}\right]$. Based on the empirical results reported above, it is reasonable to select $\left(K + 1\right) = 8$ intervals, starting with 60 observations and for convenience to end with 250 observations (one trading year), i.e., we consider the set
\vspace{-0.50cm}
\begin{center}
$\left\{60, 72, 86, 104, 124, 149, 179, 250\right\}$.
\end{center}
\vspace{-0.50cm}
We assume that the model parameters are constant within the initial interval $I_{0}$. Furthermore, $c = 1.20$ is selected in accordance with current literature.

\textbf{Local Change Point Detection Test}

A sequential testing procedure enables us to adaptively find the homogeneous interval at a fixed data point $t_{0}$. Assuming that $I_{0}$ is homogeneous, consider now the interval $J_{k} = I_{k} \setminus I_{k-1}$, and sequentially conduct the test, over interval index steps $k = 1,\ldots, K$. The hypotheses of the test at step $k$ read as
\vspace{-0.50cm}
\begin{center}
$H_{0}:$ parameter homogeneity of $I_{k}$ vs $H_{1}: \exists$ change point within $J_{k} = I_{k} \setminus I_{k-1}$.
\end{center}
\vspace{-0.50cm}

The test statistics is
\begin{equation}
T_{k, \tau} = \underset{s \in J_{k}}{\operatorname{sup}} \left\{\ell_{A_{k, s}}\left(\mathcal{Y}, \widetilde{\theta}_{A_{k, s}, \tau}\right) + \ell_{B_{k, s}}\left(\mathcal{Y}, \widetilde{\theta}_{B_{k, s}, \tau}\right) - \ell_{I_{k+1}}\left(\mathcal{Y}, \widetilde{\theta}_{I_{k+1}, \tau}\right) \right\}
\label{CARE_LCP}
\end{equation}
where $A_{k, s} = \left[t_{0}-n_{k+1},s\right]$ and $B_{k, s} = \left(s, t_{0}\right]$ are subintervals of $I_{k+1}$. Since the change point position is unknown, we test every point $s \in J_{k}$.

\begin{figure}
\begin{center}
\includegraphics[scale = 0.85]{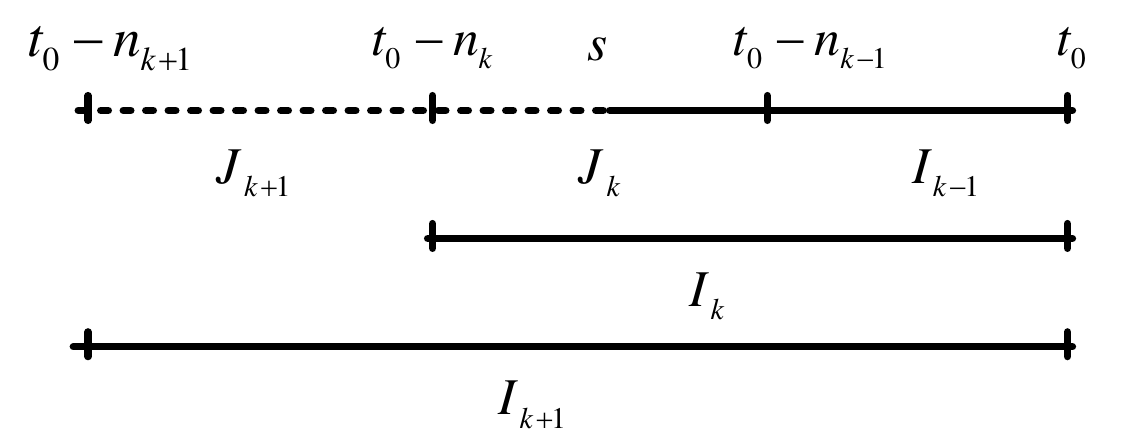}
\caption{Sequential testing for parameter homogeneity in interval $I_{k}$ with length $n_{k}$ ending at fixed time point $t_{0}$}
\label{CARE_Figure3}
\end{center}
\end{figure}

The algorithm at step $k$ is visualized in Figure \ref{CARE_Figure3}. Assuming that the null of homogeneity of interval $I_{k - 1}$ has not been rejected, the testing procedure at step $k$ tests for the homogeneity of $I_{k}$.  Since the position of a change point within $J_{k} = I_{k} \setminus I_{k - 1}$ is unknown, the test statistic is calculated based on all points $s \in J_{k}$, i.e. $s \in \left(t_{0}-n_{k-1},t_{0}-n_{k}\right]$, utilizing data from $I_{k + 1}$. Compute the sum of the log-likelihood values over the sample interval $A_{k, s} = \left[t_{0}-n_{k+1},s\right]$ (dotted area) and $B_{k, s} = \left(s, t_{0}\right]$ (solid area) and subtract the log-likelihood value over $I_{k + 1}$. The likelihood ratio test statistics $T_{k, \tau}$ at each predetermined expectile level $\tau$ is then found by \eqref{CARE_LCP}.

%The distribution of the test statistic is unknown and we therefore need to simulate critical values for each modelling setup.

In order to identify the homogeneous interval length, the test statistic \eqref{CARE_LCP} is at every step $k = 1, \ldots, K$ compared to the corresponding simulated critical value, here denoted by $\mathfrak{z}_{k, \tau}$ and elaborated below. If the test statistics at all steps up to and including $k$ are lower than the critical values, we do not reject the null hypothesis that $I_{k}$ is homogeneous and proceed to the next $\left(k + 1\right)$-st step. If, however, the test statistic firstly exceeds the critical value at step $k$, $I_{k - 1}$ is our adaptive choice. For convenience, we denote by $\widehat{k}$ the index of the interval of homogeneity. If the null is already rejected at the interval $I_{1}$, $\widehat{k} = 0$ and similarly, if $I_{K}$ represents the interval of homogeneity, $\widehat{k} = K$.

%The critical value sequence essentially controls the threshold values of the likelihood ratio test statistic.

The adaptive estimate is finally represented by the QMLE at the interval of homogeneity. Formally, it is obtained by $\widehat{\theta}_{\tau} = \widetilde{\theta}_{I_{\hat{k}}, \tau}$, with $\widehat{k} = \underset{k \leq K}{\operatorname{max}}\left\{k: T_{\ell, \tau} \leq \mathfrak{z}_{\ell, \tau}, \ell \leq k \right\}$. Here the index and the length of the interval of homogeneity are denoted by $\widehat{k}$ and $n_{\widehat{k}}$, respectively. Again, if the null is already rejected at the interval $I_{1}$, $\widehat{\theta}_{\tau} = \widetilde{\theta}_{I_{0}, \tau}$ and if $I_{K}$ is selected, $\widehat{\theta}_{\tau} = \widetilde{\theta}_{I_{K}, \tau}$. Before presenting our key empirical results we now discuss the basic idea of calculating critical values and provide at the end of the chapter a summary of the LCP testing procedure.

\textbf{Critical Values}

The critical value defines the level of significance for the aforementioned test statistic \eqref{CARE_LCP}. In classical hypothesis testing, critical values are selected to ensure a prescribed test level, the probability of rejecting the null under the null hypothesis (type I error). In the considered framework, we similarly control the loss of this 'false alarm' of detecting a non-existing change point.

Under the null hypothesis of time-invariant parameters, the desired interval of homogeneity is the longest interval $I_{K}$. When the selected interval is relatively shorter, one effectively detects a non-existing change point, here regarded as a 'false alarm'. Therefore, we aim controlling the loss associated with selecting the adaptive estimate $\widehat{\theta}_{\tau} = \widetilde{\theta}_{I_{\widehat{k}}, \tau}$ instead of $\widetilde{\theta}_{I_{K}, \tau}$: the loss is stochastically bounded
\begin{equation}
\operatorname{\E}_{\theta^{\ast}_{\tau}}\left|\ell_{I_{K}}\left(\mathcal{Y}; \widetilde{\theta}_{I_{K}, \tau}\right) - \ell_{I_{K}}\left(\mathcal{Y}; \widehat{\theta}_{\tau}\right)\right|^{r} \leq \rho\mathcal{R}_{r}\left(\theta^{\ast}_{\tau}\right)
\label{CARE_condition}
\end{equation}
where $\rho$ denotes a given significance level, see, e.g., \citet{Spokoiny:09}. This condition \eqref{CARE_condition} ensures that the loss associated with 'false alarm' (i.e., selecting $\widehat{k} < K$) is at most equal to a $\rho-$fraction of the parametric risk bound \eqref{CARE_KL}.

In a similar way at each step $k = 1, \ldots, K$, the algorithm satisfies the so-called propagation condition
\begin{equation}
\operatorname{\E}_{\theta^{\ast}_{\tau}}\left|\ell_{I_{k}}\left(\mathcal{Y}; \widetilde{\theta}_{I_{k}, \tau}\right) - \ell_{I_{k}}\left(\mathcal{Y}; \widehat{\theta}_{\tau}\right)\right|^{r} \leq \rho_{k}\mathcal{R}_{r}\left(\theta^{\ast}_{\tau}\right)
\label{CARE_propagation_condition}
\end{equation}
with $\rho_{k} = \displaystyle\frac{\rho k}{K}$ and the adaptive estimator $\widehat{\theta}_{\tau} = \widetilde{\theta}_{I_{k}, \tau}$. This propagation condition \eqref{CARE_propagation_condition} controls not only the frequency but also accounts for the deviation of the selected (adaptive) estimate from the unknown 'true' parameter. A relatively small likelihood loss on the left hand side of equation \eqref{CARE_propagation_condition} implies that the adaptive estimate $\widehat{\theta}_{\tau}$ lies with high probability in the confidence set of the optimal parameter $\widetilde{\theta}_{I_{k}, \tau}$ within the interval $I_{k}$. A large deviation value indicates that the adaptive estimate belongs to the confidence set of the optimal estimate with a small probability, i.e., $\widehat{\theta}_{\tau}$ differs significantly from $\widetilde{\theta}_{I_{k}, \tau}$ and there may be a change point presented within the interval  $I_{k}$. Under homogeneity at every step up to $k$, it is ensured that the adaptive selected homogenous interval $I_{\widehat{k}}$ extends to the underlying optimal $I_{k}$ with high probability.

%Further, the choice of the tuning parameters $r$ and the probability level $\rho$ depends on the applications at the hand.
The power loss $r$ close to zero ($r \rightarrow 0$) leads back to only counting the occurrence of false alarms. Larger risk power levels also account for the deviation of the adaptive estimate to the true parameter. Equation \eqref{CARE_propagation_condition} provides the essential requirements of calculating critical values. A detailed description of the simulation steps, as well as the resulting critical values figures are for convenience provided in Appendix \ref{Appendix_02}.

\textbf{LCP Detection Test in Practice}

The scheme of the conducted LCP detection test at fixed time point $t_{0}$, expectile level $\tau$, risk power $r$ and $\rho$ is:
\begin{enumerate}
	\item Select intervals $I_{k + 1}$, $J_{k}$, $A_{k, s}$ and $B_{k, s}$ at step $k$ and compute the test statistics $T_{k, \tau}$, see equation \eqref{CARE_LCP}
	\item Testing procedure - select the set of critical values according to the persistence parameter estimate $\widetilde{\alpha}_{1}$ (based on $I_{K}$), see Appendix \ref{Appendix_02}
	\item Interval of homogeneity - interval $I_{\widehat{k}}$ for which the null has been first rejected at step $\widehat{k} + 1$; $\widehat{k} = \underset{k \leq K}{\operatorname{max}}\left\{k: T_{\ell, \tau} \leq \mathfrak{z}_{\ell, \tau}, \ell \leq k \right\}$
	\item Adaptive estimation - the adaptively estimated parameter vector equals the QMLE at the interval of homogeneity $\widehat{\theta}_{\tau} = \widetilde{\theta}_{I_{\hat{k}}, \tau}$.
\end{enumerate}

\section{Empirical Results}
\label{EMP}

lCARE accommodates and reacts to structural changes. From the fixed rolling window exercise in subsection \ref{PD} one observes time-varying parameter characteristics while facing the trade-off between parameter variability and the modelling bias. How to account for the effects of potential market changes on the tail risk based on the intervals of homogeneity? In this section, we utilize the lCARE model to estimate the tail risk exposure across three stock markets. Using the time series of the adaptively selected interval length, we improve a portfolio insurance strategy employing our tail risk estimate and furthermore enhance its performance in the financial applications part.

\subsection{Intervals of Homogeneity}	

\begin{figure}
\begin{center}
\includegraphics[scale = 0.6]{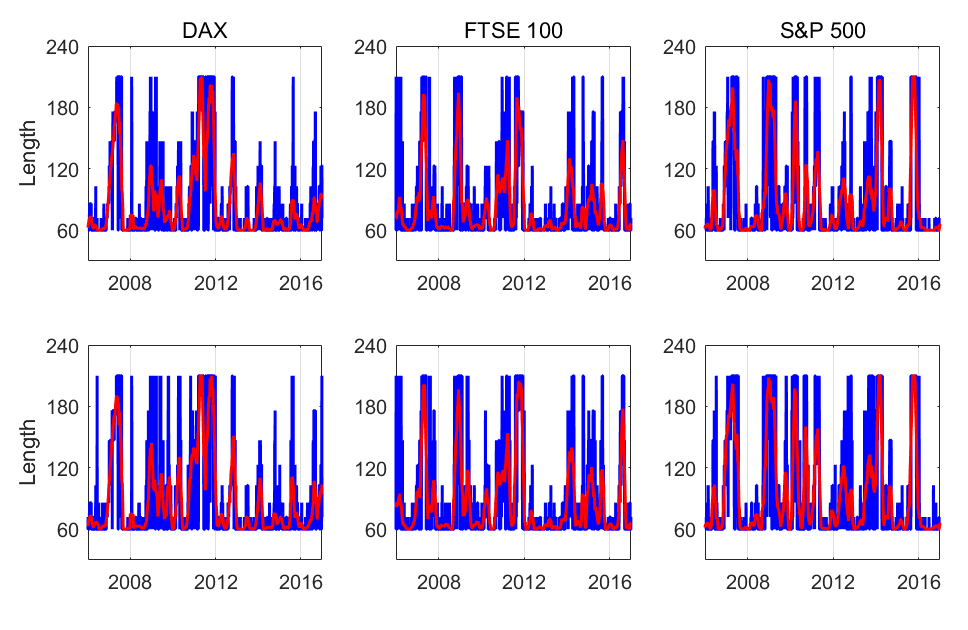}
\caption{Estimated length of the interval of homogeneity in trading days across the selected three stock markets from 2 January 2006 to 30 December 2016 for the modest (upper panel, $r = 0.8$) and the conservative (lower panel, $r = 1$) risk cases. The expectile level equals $\tau = 0.05$.
%\hspace*{\fill} \raisebox{-1pt}{\includegraphics[scale=0.05]{qletlogo}}
%\href{https://github.com/QuantLet/lCARE-BTU-HUB/tree/master/LCARE_Adaptive_Estimation_Length_005}{\,LCARE\_Adaptive\_Estimation\_Length\_005}
%\raisebox{-1pt}{\includegraphics[scale=0.05]{qletlogo}}\href{https://github.com/QuantLet/lCARE-BTU-HUB/tree/master/LCARE_Adaptive_Estimation_005}{LCARE\_Adaptive\_Estimation\_005}
}
\label{CARE_7_adaptive_estimation_length_005}
\end{center}
\end{figure}

\begin{figure}
\begin{center}
\includegraphics[scale = 0.6]{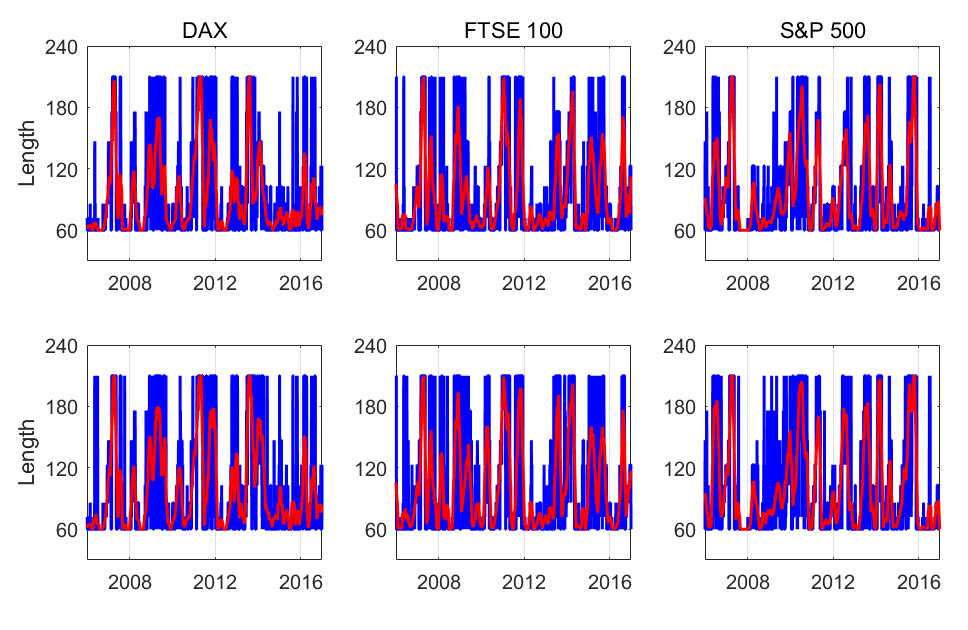}
\caption{Estimated length of the interval of homogeneity in trading days across the selected three stock markets from 2 January 2006 to 30 December 2016 for the modest (upper panel, $r = 0.8$) and the conservative (lower panel, $r = 1$) risk cases. The expectile level equals $\tau = 0.01$.
%\hspace*{\fill} \raisebox{-1pt}{\includegraphics[scale=0.05]{qletlogo}}
%\href{https://github.com/QuantLet/lCARE-BTU-HUB/tree/master/LCARE_Adaptive_Estimation_Length_001}{\,LCARE\_Adaptive\_Estimation\_Length\_001}
%\raisebox{-1pt}{\includegraphics[scale=0.05]{qletlogo}}\href{https://github.com/QuantLet/lCARE-BTU-HUB/tree/master/LCARE_Adaptive_Estimation_001}{\,LCARE\_Adaptive\_Estimation\_001}
}
\label{CARE_7_adaptive_estimation_length_001}
\end{center}
\end{figure}

\begin{figure}
\begin{center}
\includegraphics[scale = 0.6]{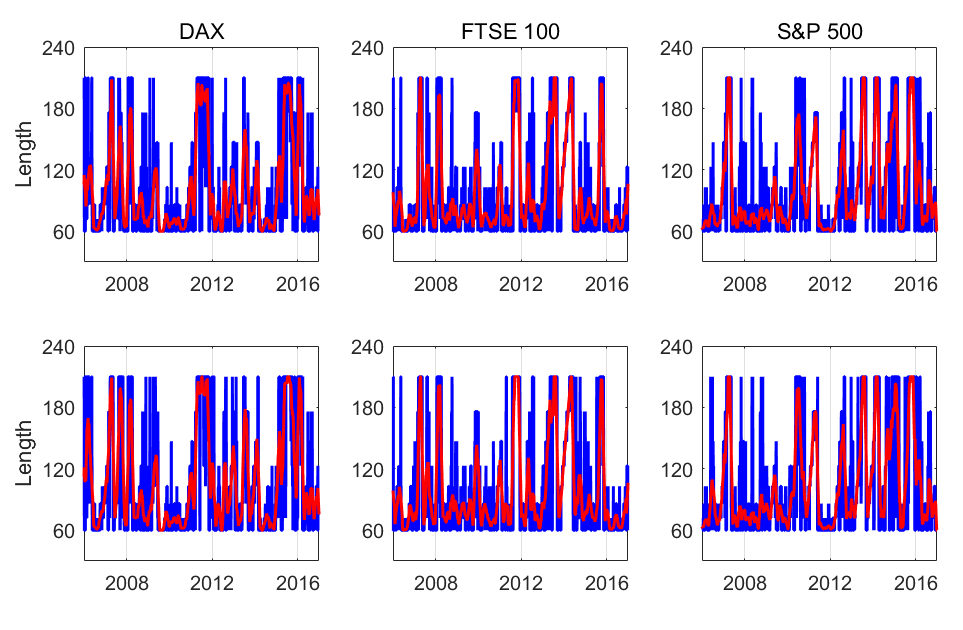}
\caption{Estimated length of the interval of homogeneity in trading days across the selected three stock markets from 2 January 2006 to 30 December 2016 for the modest (upper panel, $r = 0.8$) and the conservative (lower panel, $r = 1$) risk cases. The expectile level equals $\tau = 0.0025$.
%\hspace*{\fill} \raisebox{-1pt}{\includegraphics[scale=0.05]{qletlogo}}
%\href{https://github.com/QuantLet/lCARE-BTU-HUB/tree/master/LCARE_Adaptive_Estimation_Length_001}{\,LCARE\_Adaptive\_Estimation\_Length\_001}
%\raisebox{-1pt}{\includegraphics[scale=0.05]{qletlogo}}\href{https://github.com/QuantLet/lCARE-BTU-HUB/tree/master/LCARE_Adaptive_Estimation_001}{\,LCARE\_Adaptive\_Estimation\_001}
}
\label{CARE_7_adaptive_estimation_length_00025}
\end{center}
\end{figure}

The interval of homogeneity in tail expectile dynamics is obtained here by the lCARE framework for the time series of DAX, FTSE 100 and S\&P 500 returns. Using the sequential local change point detection test, the optimal interval length is considered at three expectile levels, namely, $\tau = 0.0025$, $\tau = 0.01$ and $\tau = 0.05$. The homogeneity intervals are interestingly relatively longer at the end of 2009 and at the beginning of 2010, especially at $\tau = 0.05$, the period following the financial crisis across all three stock markets, see, e.g., Figures \ref{CARE_7_adaptive_estimation_length_005}, \ref{CARE_7_adaptive_estimation_length_001} and \ref{CARE_7_adaptive_estimation_length_00025}. All figures present the estimated lengths of the interval of homogeneity in trading days across the selected three stock market indices from 2 January 2006 to 30 December 2016. The upper panel depicts the modest risk case $r = 0.5$, whereas the lower panel denotes the conservative risk case $r = 1$.

Recall that the lCARE model selects the longest interval over which the null hypothesis of time homogeneity of CARE parameters is not rejected. In the financial crisis initial period, the homogeneity intervals became shorter, due to the increasing market volatility and obvious market turmoil. During the post-crisis period, characterised by the high volatile regime, the homogeneity intervals became relatively longer.

\begin{table}[htp]
\begin{center}
\begin{tabular}{c|ccc|ccc}
\hline\hline
 & \multicolumn{3}{c}{$ r = 0.8 $} \vline & \multicolumn{3}{c}{$ r = 1.0 $} \\
 & DAX & FTSE 100 & S\&P 500   & DAX & FTSE 100 & S\&P 500 \\
\hline
 $\tau = 0.05$ &     85  &     83  &     90  &     89  &     88  &     95  \\
 $\tau = 0.01$ &     93  &     95  &     95  &    100  &    102  &    101  \\
 $\tau = 0.0025$ &    101  &     97  &    101  &    108  &     99  &    104  \\
\hline
\hline
\end{tabular}
\caption{Mean value of the adaptively selected intervals. Note: the average number of trading days of the adaptive interval length is provided for the DAX, FTSE 100 and S\&P 500 market indices at three expectile levels, $\tau = 0.05$, $\tau = 0.01$ and $\tau = 0.0025$, and the modest $\left(r = 0.50\right)$ and the conservative $\left(r = 1.00\right)$ risk case.}
\label{CARE_adaptive_interval}
\end{center}
\end{table}
\vspace{-0.50cm}

In a similar way, the intervals of homogeneity are relatively shorter in the modest risk case $r = 0.8$, as compared to the conservative risk case $r = 1$. The average daily selected optimal interval length supports this, see, e.g., Table \ref{CARE_adaptive_interval}. The results are presented for all expectile levels at the modest and the conservative risk cases, $r = 0.80$ and $r = 1$, respectively. At expectile levels $\tau = 0.0025$ and $\tau = 0.01$, the intervals of homogeneity are slightly larger than the intervals at $\tau = 0.05$.

\subsection{Dynamic Tail Risk Exposure}

\begin{figure}
\begin{center}
\includegraphics[scale = 0.6]{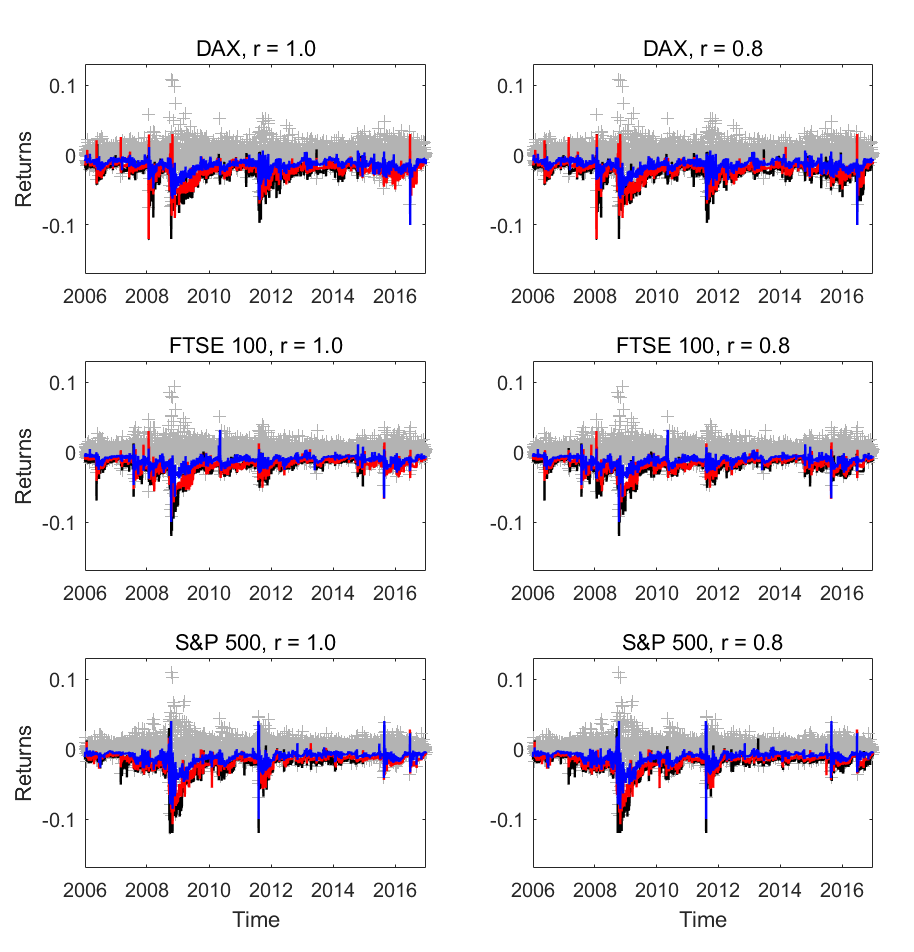}
\caption{Estimated expectile risk exposure at level $\tau = 0.05$ (blue), $\tau = 0.01$ (red) and $\tau = 0.0025$ (black) for return time series of DAX, FTSE 100, and S\&P 500 indices from 2 January 2006 to 30 December 2016. The left panel shows results of the conservative risk case $r = 1$ and the right panel depicts the results of the modest risk case $r = 0.8$. }
\label{CARE_expectile}
\end{center}
\end{figure}

Based on the lCARE model, one can directly estimate dynamic tail risk exposure measures using the adaptively selected intervals. The tail risk at smaller expectile level is lower than risk at higher levels, see, e.g., Figure \ref{CARE_expectile}. Here the estimated expectile risk exposure for the three stock market indices from 2 January 2006 to 30 December 2016 is displayed for all three expectile levels. The left panel represents the conservative risk case $r = 1$ results, whereas the right panel considers the modest risk case $r = 0.8$. The former leads on average to slightly lower variability, as compared to the modest risk which results in shorter homogeneity intervals.

%% 3
%\newpage \newgeometry{left = 2.5cm, right = 2.5cm, top = 2.5cm, bottom = 2.5cm}
%\begin{landscape}
%\begin{figure}
%\begin{center}
%\includegraphics[scale = 1.0]{Figures/lCARE_expectile_figure_rote.png}
%\caption{Estimated expectile risk exposure at level $\tau = 0.05$ (blue), $\tau = 0.01$ (red) and $\tau = 0.0025$ (black) for return time series of DAX, FTSE 100, and S\&P 500 indices from 2 January 2006 to 30 December 2016. The left panel shows results of the conservative risk case $r = 1$ and the right panel depicts the results of the modest risk case $r = 0.8$. }
%\label{CARE_expectile}
%\end{center}
%\end{figure}
%\end{landscape}
%\restoregeometry

The estimated expectiles allow us to compute other tail risk measures, most prominently expected shortfall - the expected value of portfolio loss above a certain threshold, e.g., Value at Risk (VaR). The quantile estimation can be improved by employing an expectile-based expected shortfall (ES) framework. In its derivation one notes a one-to-one mapping between quantiles and expectiles with the expectile level $\tau_{\alpha}$ being selected such that $e_{t, \tau_{\alpha}} = q_{\alpha}$, i.e., $\alpha$-quantile
\begin{equation}
\tau_{\alpha} \ = \displaystyle\frac{ \alpha \cdot q_{\alpha} - \displaystyle\int^{q_\alpha}_{-\infty}ydF(y) }{ \E\left[Y\right] - 2 \displaystyle\int^{q_\alpha}_{-\infty}ydF(y) - \left(  1 -  2\alpha \right) q_{\alpha}}
\label{LCARE_tau_alpha}
\end{equation}
where $F\left(\cdot\right)$ denotes the cumulative density function (cdf) of a random variable $Y$. The corresponding expected shortfall can be expressed as
\begin{equation}
	ES_{e_{t, \tau_{\alpha}}} = \left| 1 + \tau_{\alpha}\left(1 - 2\tau_{\alpha}\right)^{-1}\alpha^{-1} \right|e_{t, \tau_{\alpha}}
\label{LCARE_ES}
\end{equation}
\vspace{-0.20cm}
with $e_{t, \tau_{\alpha}}$ denoting the expectile at level $\tau_{\alpha}$. In order to apply \eqref{LCARE_ES}, one needs to fix a certain cdf $F\left(\cdot\right)$ in \eqref{LCARE_tau_alpha}. For convenience we chose the asymmetric normal distribution.

\begin{figure}[htp]
\begin{center}
\includegraphics[scale = 0.6]{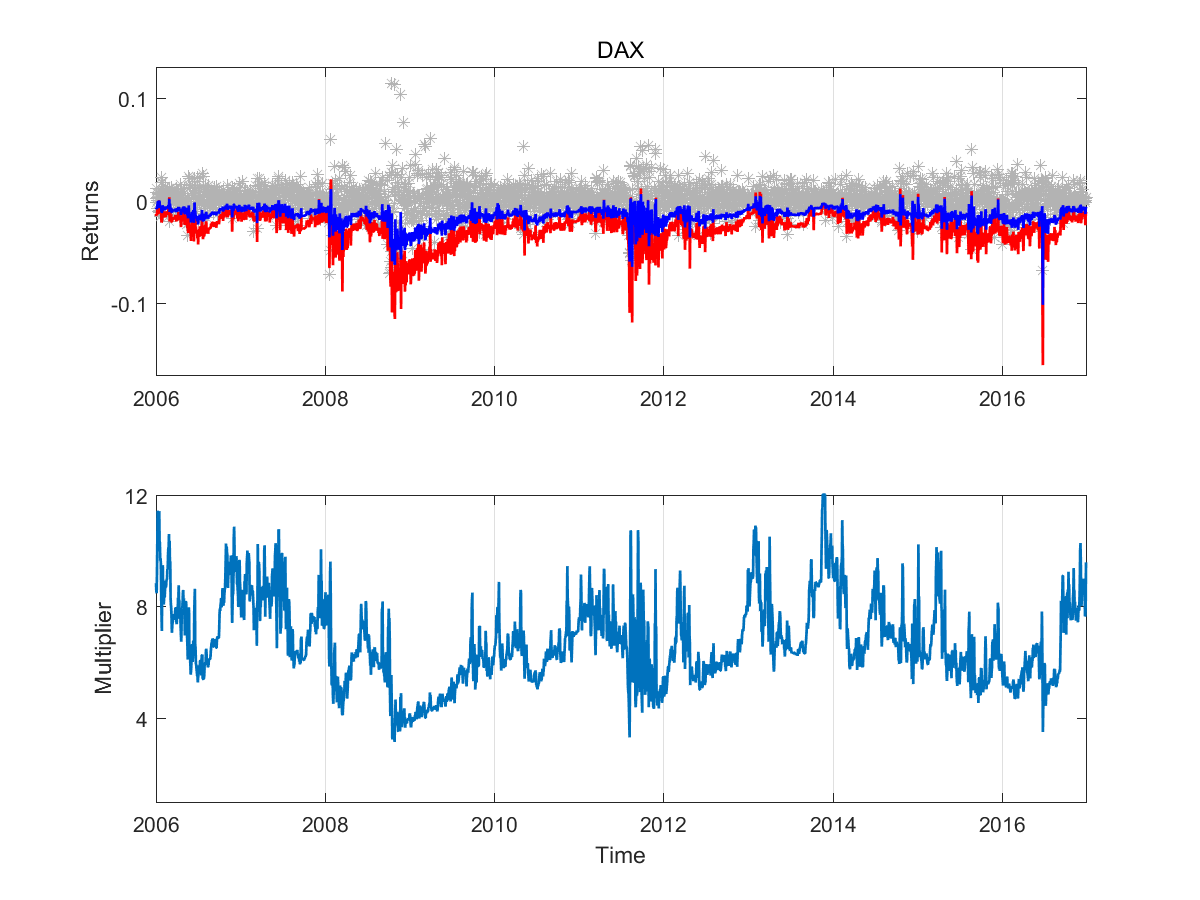}
% \vspace{0.20cm}
\caption{Adaptively estimated expectile (blue) and expected shortfall (red) series for DAX index returns from 2 January 2006 to 30 December 2016 (upper panel). The lower panel shows the corresponding multiplier dynamics. We choose $r = 1$ and $\tau = 0.05$.}
\label{CARE_DAX_ES}
\end{center}
\end{figure}

\begin{figure}[htp]
\begin{center}
\includegraphics[scale = 0.6]{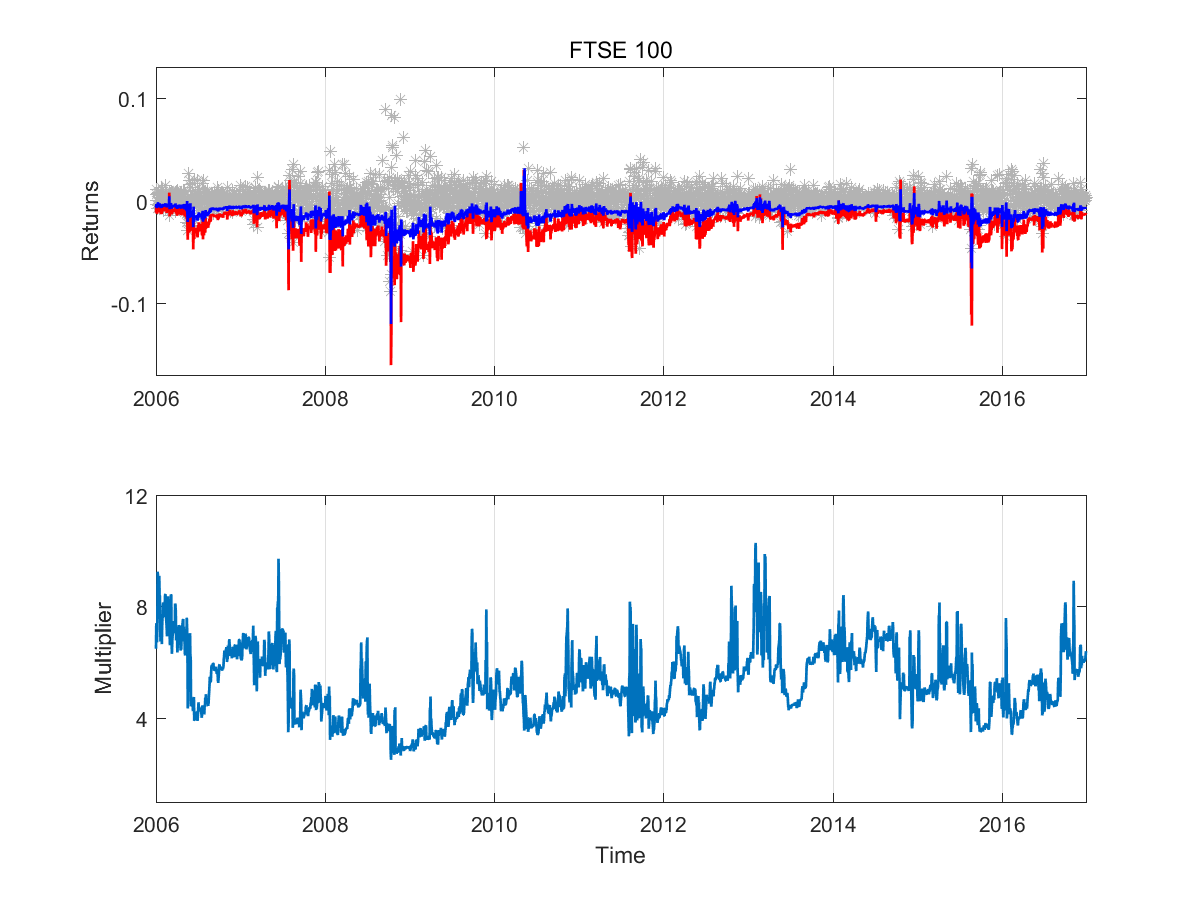}
% \vspace{0.20cm}
\caption{Adaptively estimated expectile (blue) and expected shortfall (red) series for FTSE 100 index returns from 2 January 2006 to 30 December 2016 (upper panel). The lower panel shows the corresponding multiplier dynamics. We choose $r = 1$ and $\tau = 0.05$.}
\label{CARE_FTSE_ES}
\end{center}
\end{figure}

\begin{figure}[htp]
\begin{center}
\includegraphics[scale = 0.6]{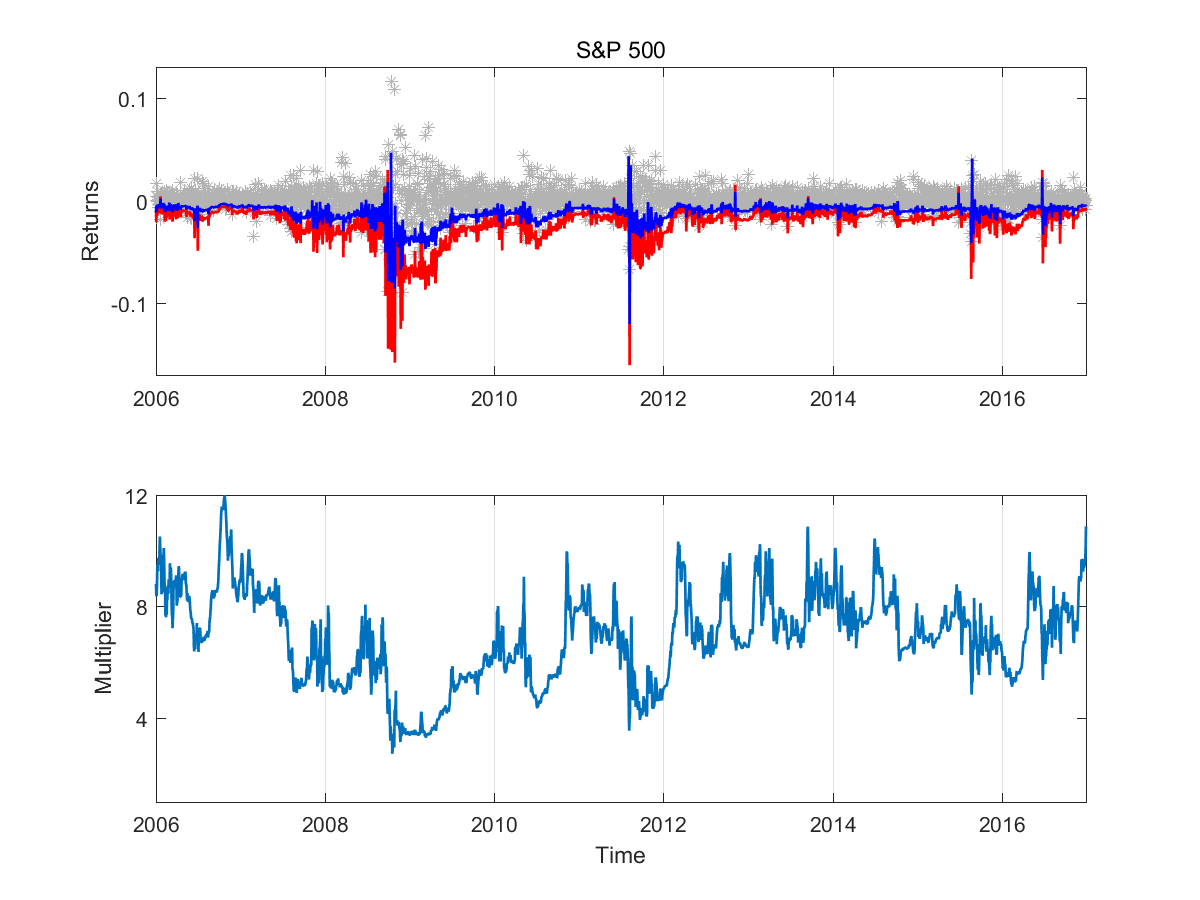}
% \vspace{0.20cm}
\caption{Adaptively estimated expectile (blue) and expected shortfall (red) series for S\&P 500 index returns from 2 January 2006 to 30 December 2016 (upper panel). The lower panel shows the corresponding multiplier dynamics. We choose $r = 1$ and $\tau = 0.05$.}
\label{CARE_SP_ES}
\end{center}
\end{figure}

Consider the tail risk exposure of DAX, FTSE 100 and S\&P 500 index series at expectile level $\tau = 0.05$ and conservative risk case $r = 1.0$. During market distress periods, the 2008 financial crisis and the 2012 European sovereign debt crisis, the estimated expected shortfall \eqref{LCARE_ES} exhibits a high variation as depicted in the upper panel of figures \ref{CARE_DAX_ES}, \ref{CARE_FTSE_ES}, \ref{CARE_SP_ES}. Similarly to current research developments, the estimated expected shortfall using the proposed lCARE model exceeds (by magnitude) the estimated expectile $e_{t, \tau}$ value.

\subsection{Application: Portfolio Insurance}

Dynamic tail risk measures are useful tools in quantitative practice. Portfolio insurance deals, for instance, with (portfolio) protection strategies tailored especially for mutual fund management while solving portfolio optimization tasks. Consider particularly the task of preserving a given proportion of an initial asset portfolio value at the end of the predetermined time horizon. In this strategy the downside risk is limited under bearish market conditions and simultaneously the optimal profit return emerges in bullish market situations and thus fund managers can utilize the time invariant portfolio protection (TIPP), \citet{Estep:88}, \citet{Hamidi:14}. It turns out that this represents an extension of the constant proportion portfolio insurance (CPPI) strategy by \citet{Black:87}, \citet{Black:92}.

In practice, the fund managers firstly determine a floor - a lowest acceptable portfolio value at the end of the investment horizon. Then the exposure, the multiple amount of the excess of the portfolio value above the floor by a multiplier, is invested into the risky and the remaining part into a riskless asset. The underlying floor of the TIPP strategy is time-varying as compared to the CPPI method. In this aspect, the floor is related to a proportion of the highest previous portfolio value, which seems more conservative, however, more actively responds to the prevailing market conditions.

The proportion of the total portfolio value invested in risky assets is determined by the so-called asset multiplier representing the leverage value of the risky exposure. A traditional approach assumes that the multiplier is a constant, i.e., insensitive to the current market conditions. Our lCARE model adapts to the current risk exposure at different states of the economy (bearish or bullish market), since we account for the time-varying properties of the asset multiplier in portfolio allocation. It is expected that during favourable conditions, more wealth can be allocated into risky investments and vice versa. In this part, the trading idea of the TIPP strategy is explained and thereafter the relationship between the multiplier and the return of the risky asset is presented. The methodologies (constant vs adaptive multiplier selection) are then applied to the DAX, FTSE 100 and S\&P 500 series and evaluated afterwards.

\textbf{Time Invariant Portfolio Protection Strategy (TIPP)}

Denote the initial asset portfolio value as $V_{t}$ at time $t \in (0, T]$. An investor aims to preserve a predetermined protection value $F_{t}^{s}$, the so-called floor, at each day
\begin{equation}	
V_t \geq  \ s \times {\operatorname{max}}  \left\{F \cdot e^{-rf_{t} \cdot (T-t)}, \  \underset{p\leq t}{\operatorname{sup}} \ V_p  \right\} = F_{t}^{s}
\label{CARE_TIPP_portfolio_value}
\end{equation}
with an exogenous parameter $s \in (0, 1)$ and the cushion value, $C_t = V_t - F_{t}^{s} \geq 0$. $rf_{t}$ is the risk-free rate, we set the initial value $F = 100$ and the proportion value $s = 0.9$. The allocation decision states that $G_t = m \cdot C_t$ is invested into the risky asset with return $r_{t}$ (here the index portfolio) where $m$ denotes a non-negative multiplier that controls the portfolio performance. The remaining amount $V_t - G_t$ is invested into a riskless asset.

The portfolio value $V_t$ and consequently the cushion value $C_t = V_t - F_{t}^{s}$ evolve as
\vspace{-0.25cm}
\begin{align}
V_{t + 1} &= V_{t} \ + G_{t} r_{t+1}\ + \left(V_{t} - G_{t} \right)rf_{t + 1} \\
C_{t + 1} &= C_{t} \left \{ 1\ + m \cdot r_{t+1} \ + \left(1 - m\right)rf_{t + 1} \right \}
\label{CARE_TIPP_cushion}
\end{align}
Since the cushion value $C_t \geq 0$, for all $t \leq T$, an upper bound of the multiple $m$ can be derived from equation \eqref{CARE_TIPP_cushion} when $rf_{t}$ is negligibly small and the risky asset return is negative
\vspace{-0.3cm}
\begin{equation}
     m \leq \left(- r^{-}_{t + 1} \right)^{-1}, \ \forall t \leq T
\label{CARE_TIPP_multiplier}
\vspace{-0.3cm}
\end{equation}
with $r^{-}_{t + 1} = \min (0, r_{t + 1})$.

This equation \eqref{CARE_TIPP_multiplier} reflects a relationship between $m$ and the tail structure of the distribution of $r_{t}$. When the downside return loss is, for example, 10\%, $m \leq 10$, and for a downside of 20\%, $m \leq 5$. When the market is bullish (bearish), the investor is more prone to invest into the risky (risk-free) asset.

In the above TIPP strategy, the cushion value is always expected to be near or above zero. This property only holds in continuous time and assumes that the investor could timely modify their portfolio allocation before a large downside return happens. In practice, fund managers have to account for the risk that the cushion value may be negative since there may happen a unpredictable large downside market movement whereupon the managers may fail to reschedule their portfolio allocations in the discontinuous rebalancing. This risk is known as the gap risk.

How to deal with gap risk and correspondingly calculate the multiplier? There are two common approaches: the first is through the quantile hedging method, see e.g. \citet{Follmer:99}, exploiting VaR to imply the multiplier; another method is based on expected shortfall, see e.g. \citet{Hamidi:14}, \citet{Ameur:14}. In the quantile hedging framework, for a given level $\alpha$, the protection portfolio condition is given by
\vspace{-0.50cm}
\begin{center}
$\PP\left(C_t \geq 0, \ \forall t \leq T\right) \geq 1 - \alpha$.
\end{center}
\vspace{-0.50cm}
Similar to the derivation of \eqref{CARE_TIPP_multiplier}, the multiplier can now be expressed as the $\left(1-\alpha\right)$-th quantile of the return distribution
\begin{equation*}
	\PP  \left\{ m_{t} \leq \left(- r^{-}_{t + 1} \right)^{-1} , \ \forall t \leq T \right\} \ \geq \ 1 - \alpha
\end{equation*}
where the bound of $m$ with quantile can be obtained by the above equation.

The expected shortfall is a coherent risk measure and is more suitable to reflect the tail risk since the quantile technique does not take the magnitude of tail risk at all into account.  When the investor is prone to more conservative asset allocation, ES is proposed to estimate the multiplier, see \citet{Hamidi:14}.

\textbf{Performance Comparison}

Here we employ the lCARE method to estimate ES controlling the gap risk. The corresponding multiplier selection is thus expressed by the lCARE-based ES
\begin{equation}
	m_{t, \tau} = \left| ES_{e_{t, \tau}}\right|^{-1}
\end{equation}
with $e_{t, \tau}$ denoting the associated expectile value. The conditional multiplier is the inverse of the expected shortfall. In practice, we assume that the data process follows an asymmetric normal distribution, and the threshold range for $m_{t, \tau} \in \left\{1, 2, \ldots 12\right\}$ is used. The dynamics of the implied multipliers for the selected indices corresponding to ES estimates are displayed in the lower panel of figures \ref{CARE_DAX_ES}, \ref{CARE_FTSE_ES}, \ref{CARE_SP_ES}, based on the lCARE model with $r = 1$ and $\tau = 0.05$ from 2 January 2006 to 30 December 2016 for the DAX, FTSE 100 and S\&P 500 series, respectively.

The one-year rolling window estimation strategy is also selected as one of the benchmark models. In the appendix, the left panel of figures \ref{CARE_DAX_TIPP_rolling}, \ref{CARE_FTSE_TIPP_rolling}, \ref{CARE_SP_TIPP_rolling} presents the estimated expectile and ES based on a one-year fixed rolling window estimation and the corresponding multipliers for the three stock markets respectively. The constant multiplier cases (from 1 to 12) are included for benchmark comparisons as well.

ES can also be implied by the CAViaR framework, one of the popular conditional autoregressive modelling approaches for the Value at Risk. Given a one-to-one mapping between expectiles and quantiles, the expected shortfall can be formulated by the quantile at the corresponding quantile level when the expectile and quantile values are equal, see \eqref{LCARE_tau_alpha}. Here we include the CAViaR based ES as another benchmark and provide its corresponding multiplier dynamics that are implemented in the insurance strategy. We firstly choose the corresponding quantile level, then illustrate the CAViaR specification from \citet{Engle:04}, before presenting the final results.

Under the asymmetric normal distribution assumption, given expectile level $\tau = 0.05$, Equation \eqref{LCARE_tau_alpha} implies the corresponding quantile level $\alpha = 0.065$. While \citet{Engle:04} state four CAViaR model specifications, the following model specification, similar to equation \eqref{LCARE_Expectile_Specification}, is selected
\vspace{-0.35cm}
\begin{align}
y_{t} &= q_{t, \alpha} + \varepsilon_{t, \alpha} \quad\; Quant_{\alpha}(\varepsilon_{t, \alpha} | \mathcal{F}_{t - 1}) = 0 \\
q_{t, \alpha}  &= \beta_{0} + \beta_{1}q_{t - 1, \alpha} + \beta_{2}q_{t - 2, \alpha} + \beta_{3}q_{t - 3, \alpha} + \beta_{4}y^{+}_{t - 1} + \beta_{5}y^{-}_{t - 1}
\label{CARE_Model_CARiaR_rolling}
\end{align}
where $q_{t, \alpha}$ represents the quantile (VaR) at $ \alpha \in \left(0, 1\right)$, and $Quant_{\alpha}(\varepsilon_{t, \alpha}| \mathcal{F}_{t - 1}) $ is the $\alpha$-quantile of $\varepsilon_{t, \alpha}$ conditional on the information set $\mathcal{F}_{t - 1}$. In addition, we choose $\alpha = 0.065$ such that $e_{\tau_{\alpha}} = q_{\alpha}$ when $\tau_{\alpha} = 0.05$.

The estimated quantiles, expectiles and ES based on a one-year rolling window estimation associated to the above mentioned CAViaR model \eqref{CARE_Model_CARiaR_rolling}, with ES implied from equation \eqref{LCARE_ES}, as well as the corresponding multipliers are presented for the DAX, FTSE 100 and S\&P 500 series in the right panel of Figures \ref{CARE_DAX_TIPP_rolling}, \ref{CARE_FTSE_TIPP_rolling} and \ref{CARE_SP_TIPP_rolling} in the appendix, respectively.

%\begin{figure}
%\begin{center}
%\includegraphics[scale = 0.70]{Figures3/DAX_CAViaR_rolling_ES_multiple.pdf}
%\caption{}
%\label{CARE_DAX_CAViaR_rolling}
%\end{center}
%\end{figure}

\begin{figure}
\begin{center}
\includegraphics[scale = 0.65]{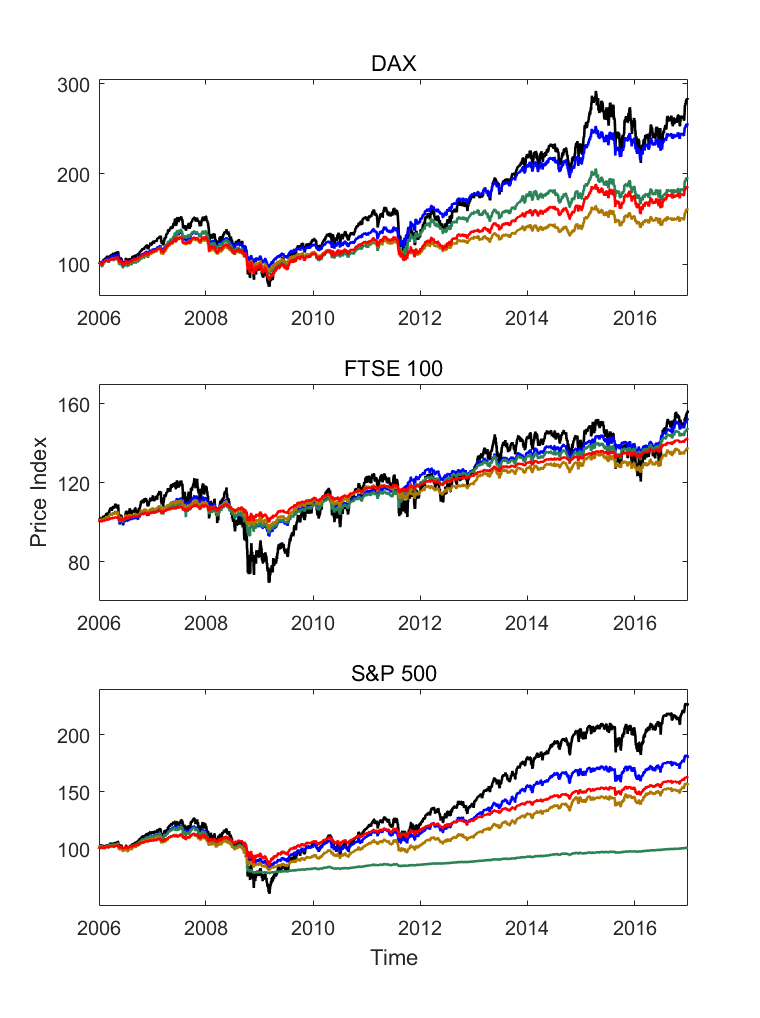}
% \vspace{0.20cm}
\caption{Performance of the portfolio value: (a) DAX index (black), (b) $m = 5$ (red), (c) one-year rolling approach (green), (d) CAViaR based one-year rolling approach ($\alpha = 0.065$) (brown), (e) $m_{t, \tau}$ - lCARE (blue) from 2 January 2006 to 30 December 2016.}
\label{CARE_performance_comparison}
\end{center}
\end{figure}

\begin{table}[hp]
\begin{center}
\begin{tabular}{c|r@{.}lr@{.}lr@{.}lr@{.}lr@{.}lr@{.}l}
\hline\hline
\multicolumn{13}{c}{Panel A     DAX} \\
\hline
 & \multicolumn{2}{c}{Return(\%)} & \multicolumn{2}{c}{Volatility(\%)} & \multicolumn{2}{c}{VaR 99\%} &
 \multicolumn{2}{c}{Skewness} & \multicolumn{2}{c}{Kurtosis} & \multicolumn{2}{c}{Sharpe} \\
\hline
DAX                  &     9&05  &   22&41  &   -4&07  &    0&14  &    9&39  &    0&03   \\
lCARE                &     8&15  &   13&09  &   -2&10  &    0&15  &    6&92  &    0&04   \\
Expectile: one-year  &     5&83  &   13&75  &   -2&30  &   -0&03  &    6&66  &    0&03   \\
CAViaR: one-year     &     4&14  &   11&13  &   -1&87  &   -0&29  &    5&10  &    0&02   \\
Multiplier 5         &     5&39  &   13&39  &   -2&51  &    0&01  &    8&94  &    0&03   \\
\hline

\multicolumn{13}{c}{Panel B     FTSE 100} \\
\hline
 & \multicolumn{2}{c}{Return(\%)} & \multicolumn{2}{c}{Volatility(\%)} & \multicolumn{2}{c}{VaR 99\%} &
 \multicolumn{2}{c}{Skewness} & \multicolumn{2}{c}{Kurtosis} & \multicolumn{2}{c}{Sharpe} \\
\hline
FTSE 100             &     3&91  &   19&21  &   -3&25  &    0&04  &   10&80  &    0&01   \\
lCARE                &     3&69  &    7&57  &   -1&30  &   -0&00  &    6&23  &    0&03   \\
Expectile: one-year  &     3&40  &    7&40  &   -1&31  &    0&11  &    9&15  &    0&03   \\
CAViaR: one-year     &     2&80  &    6&06  &   -1&04  &   -0&15  &    5&68  &    0&03   \\
Multiplier 2         &     3&09  &    3&84  &   -0&64  &   -0&00  &   10&67  &    0&05   \\
\hline

\multicolumn{13}{c}{Panel C     S\&P 500} \\
\hline
 & \multicolumn{2}{c}{Return(\%)} & \multicolumn{2}{c}{Volatility(\%)} & \multicolumn{2}{c}{VaR 99\%} &
 \multicolumn{2}{c}{Skewness} & \multicolumn{2}{c}{Kurtosis} & \multicolumn{2}{c}{Sharpe} \\
\hline
S\&P 500              &     7&08  &   19&74  &   -3&79  &   -0&09  &   14&35  &    0&02   \\
lCARE                &     5&10  &   10&79  &   -2&05  &   -0&31  &    6&21  &    0&03   \\
Expectile: one-year  &     0&08  &    6&59  &   -1&49  &   -2&62  &   41&23  &    0&00   \\
CAViaR: one-year     &     3&85  &    8&88  &   -1&53  &   -0&54  &    6&75  &    0&03   \\
Multiplier 4         &     4&20  &    7&89  &   -1&52  &   -0&23  &   14&27  &    0&03   \\
\hline
\hline
\end{tabular}
\caption{Descriptive statistics of the portfolio returns based on the TIPP strategy. We employ several models: the lCARE, one-year rolling window, CAViaR rolling window and constant muliplier approach for the DAX index, FTSE 100 and  S\&P 500 from 2 January 2006 to 30 December 2016. The investment strategy is based on a one-year investment horizon.}
\label{CARE_TIPP}
\end{center}
\end{table}
\vspace{-0.50cm}

Finally, the initial and the target value of a hypothetical portfolio at the end of one year investment horizon are both set to 100 ($F = 100$ in equation \eqref{CARE_TIPP_portfolio_value}). Associated to the cushioned portfolio strategy, the daily asset allocation decision at time $t$ is to invest the multiple amount of the difference between the portfolio value and the discounted floor up to $t$ into the stock portfolio, the rest into a riskless asset. Figure \ref{CARE_performance_comparison} presents the performance of the portfolio values based on the cushioned portfolio strategy with unconditional constant multipliers as well as the conditional time-varying multipliers. The black solid line represents the index, the blue line represents the cushioned portfolio with lCARE based conditional dynamic multiplier, the green line represents the portfolio value using a one-year fixed rolling window estimated multiplier, and the brown line presents the value under CAViaR based one-year rolling estimated multiplier. The comparatively best performed portfolio among the constant multipliers considers $m = 5$, denoted by the red line.

The cushioned portfolio with the dynamic multiplier closely tracks the observed index series and simultaneously guarantees the target portfolio value floor at the end of the investment horizon at every trading day, see Figure \ref{CARE_performance_comparison}. The lCARE strategy performs very well in comparison to the cushioned portfolio with a constant multiplier, the one-year rolling window estimation based on expectile or quantile levels.

lCARE exhibits the best return moment performance of the portfolio insurance strategy, see Table \ref{CARE_TIPP}. We list the statistical results of empirical data, the TIPP strategy with lCARE - based multiplier, one-year fixed rolling window CARE - implied multiplier, one-year rolling window CAViaR implied multiplier, and constant multipliers. The average return of lCARE based strategy, 7.36\% is larger than the counterpart based on a fixed rolling window, 5.70\%. It is also observed that the CAViaR based strategy performs less favourable. Although the lCARE strategy leads to slightly lower average returns than the observed return series of 8.79\%, it turns out that it performs favourable relative to all other benchmark strategies.

\section{Conclusions}
\label{Conclusions}

The localized conditional autoregressive expectiles (lCARE) model accounts for time-varying parameter characteristics and potential structure changes in tail risk exposure modelling. The parameter dynamics implied by a fixed rolling window exercise of three stock market indices, DAX, FTSE 100 and S\&P 500, indicates that there is a trade-off between the modelling bias and parameter variability. A local parametric approach (LPA) assumes that locally one can successfully fit a parametric model. Based on a sequential testing procedure, one determines the interval of homogeneity over which a parametric model can be approximated by a constant parameter vector.

The lCARE model adaptively estimates the tail risk exposure by relying on the (in-sample) 'optimal' interval of homogeneity. Setting the expectile levels $\tau = 0.05$ and $\tau = 0.01$, the dynamic expectile tail risk measures for the selected three stock markets are successfully obtained by lCARE. Furthermore, ES has been introduced, evaluated and employed in the asset allocation example: the portfolio protection strategy is improved by the lCARE modelling framework.

\bibliography{Literature}

\newpage

\section*{Appendix}
\appendix
\addcontentsline{toc}{section}{Appendix}
\renewcommand{\thesubsection}{\Alph{subsection}}

\subsection{Parametric Risk Bound}
\label{Appendix_01}

\textbf{Data Simulation}

Adaptive estimation of CARE parameters demands critical values as the distribution of the test statistics in our finite sample environment is unknown. Thus the proposed sequential testing procedure demands critical values that are here found by a simulation study. The training data should furthermore be obtained at each expectile level for calculating the test statistics and then simulate the corresponding critical values. This step is necessary and unavoidable. Our data used for obtaining the critical values are simulated for given expectile levels. However concerning the optimal implementation of simulation procedure, unfortunately there is almost few literature covering this issue. \citet{Gerlach:14} base on the AND assumption and develop a MCMC simulation study to estimate the autoregressive expectiles, which is published in Journal of Financial Econometrics (2015).

In the similar context, we follow \citet{Gerlach:14} and \citet{Gerlach:12} assuming the same AND framework. There are three parameters in AND, the mean, variance and scape parameters, in which the scape largely depends on the tail structure of the distribution. After initially fixing the mean and variance parameters with the empirical estimates, we set the expectile value of AND equal to the counterpart from empirical data at one specific expetile level, and then obtain the scape parameter. In this way, we can generate the independent disturbance term $\varepsilon_{t, \tau}$ in \eqref{LCARE_Expectile_Specification} at a given expectile level $\tau$.

Further, as discussed in section \ref{LCARE}, there are three pseudo true parameter constellations selected at each expectile level. These parameters are estimated from the one-year rolling sample, which is regarded as the longest homogeneous interval. % In adaptive parameter estimation we follow ...
For each pseudo true parameter vector from Table \ref{CARE_parameter_dynamics_quartiles} and for each given expectile level ($\tau = 0.0025$, $\tau = 0.01$ and $\tau = 0.05$), we have simulated 1000 sample paths using the corresponding CARE specification, and then implement the critical value calculation.

\textbf{Risk Bound}

The largest average value of the ($r$-th power) difference between the respective log-likelihood values, see equation \eqref{CARE_KL}, is taken as the corresponding risk bound. Note that the considered interval candidates in this simulation cover

\vspace{-0.50cm}
\begin{center}
 $\left\{60, 72, 86, 104, 124, 149, 179, 250\right\}$
\end{center}
\vspace{-0.50cm}
observations - see the selection details in sub-section \ref{lCARE}.

The values of the simulated risk bound $\mathcal{R}_{r}\left(\theta^{\ast}_{\tau}\right)$ across different setups are provided in Table \ref{CARE_risk_bound}. We particularly consider the modest $\left(r = 0.8\right)$ and the conservative $\left(r = 1\right)$ risk case and set three expectile levels, namely $\tau = 0.0025$, $\tau = 0.01$ as well as $\tau = 0.05$. The risk bounds are obtained by Monte Carlo simulation for each selected parameter vector corresponding to Table \ref{CARE_parameter_dynamics_quartiles} where we label the first quartile of estimated parameter values as 'low', the mean as 'mid' and the third quartile as 'high'. It turns out that the risk bounds in the conservative case are relatively larger than the bounds obtained in the modest risk case.

\begin{table}[htp]
\begin{center}
\begin{tabular}{c|ccc|ccc|ccc}
\hline\hline
%\multirow{2}{*}{Model}
 & \multicolumn{3}{c}{$\tau = 0.05$} \vline & \multicolumn{3}{c}{$\tau = 0.01$} \vline & \multicolumn{3}{c}{$\tau = 0.0025$}\\
 & Low & Mid & High   & Low & Mid & High  & Low & Mid & High\\
\hline
$ r = 0.8 $ & 0.062  &  0.034  &  0.022  &  0.041  &  0.038  &  0.023  &  0.034  &  0.042  &  0.027   \\
$ r = 1.0 $ & 0.078  &  0.047  &  0.033  &  0.054  &  0.051  &  0.035  &  0.046  &  0.054  &  0.040   \\
\hline\hline
\end{tabular}
\caption{Risk bound $\mathcal{R}_{r}\left(\theta^{\ast}_{\tau}\right)$ given three expectile levels, $\tau = 0.0025$, $\tau = 0.01$ and $\tau = 0.05$. We consider the modest $\left(r = 0.8\right)$ and the conservative $\left(r = 1.0\right)$ risk case. The risk bounds are obtained by Monte Carlo simulation for each selected parameter vector from Table \ref{CARE_parameter_dynamics_quartiles} where we label the first quartile of estimated parameters as 'low', the median as 'mid' and the third quartile as 'high'.   %\hspace*{\fill}\raisebox{-1pt}{\includegraphics[scale=0.05]{qletlogo}}\href{https://github.com/QuantLet/lCARE-BTU-HUB/tree/master/LCARE_Risk_Bound_Results}{\,LCARE\_Risk\_Bound\_Results}
}
\label{CARE_risk_bound}
\end{center}
\end{table}

\subsection{Critical Values and Adaptive Estimation}
\label{Appendix_02}

\textbf{Critical Values}

Here we present a sequential choice of critical values $\mathfrak{z}_{k, \tau}$ in practice. Considering the situation after the first $k$ steps of the algorithm, we need to distinguish between two cases: in the first, change point is detected at some step, and in the other case no change point is detected. In the first case, we denoted by $\mathcal{B}_{q}$ the event that change point is detected at step $q$,
\begin{equation}
  \mathcal{B}_{q} = \{ T_{1, \tau} \leq \mathfrak{z}_{1, \tau}, \cdots, T_{q-1, \tau} \leq \mathfrak{z}_{q-1, \tau}, T_{q, \tau} > \mathfrak{z}_{q, \tau}\}
\end{equation}
where $\widehat{\theta}_{\tau} = \widetilde{\theta}_{I_{q-1}, \tau}$ on $\mathcal{B}_{q}$, $q= 1, 2, \cdots, k$. The sequence choice of $\mathfrak{z}_{k}$ is based on the decomposition
\begin{equation}\label{propoDecom}
  \left|\ell_{I_{k}}\left(\mathcal{Y}; \widetilde{\theta}_{I_{k}, \tau}\right) - \ell_{I_{k}}\left(\mathcal{Y}; \widehat{\theta}_{\tau}\right)\right|^{r} = \displaystyle \sum ^{k}_{q = 1} \left|\ell_{I_{k}}\left(\mathcal{Y}; \widetilde{\theta}_{I_{k}, \tau}\right) - \ell_{I_{k}}\left(\mathcal{Y}; \widetilde{\theta}_{I_{q-1}, \tau}\right)\right|^{r} \; \IF(\mathcal{B}_{q})
\end{equation}
where $k \leq K $. Note that the event $\mathcal{B}_{q}$ only depends on $\mathfrak{z}_{1, \tau}, \cdots, \mathfrak{z}_{q, \tau}$. For example, $\mathcal{B}_{1}$ means $T_{1, \tau} > \mathfrak{z}_{1, \tau}$ and $\widehat{\theta}_{\tau} = \widetilde{\theta}_{I_{0}, \tau} $ for all $\widehat{k} \geq 1$. We select $\mathfrak{z}_{1, \tau}$ as the minimal value that ensures
\begin{equation}
\label{critical1}
\displaystyle \max_{k = 1, \cdots, K} \operatorname{E}_{\theta^{\ast}_{\tau}}\left|\ell_{I_{k}}\left(\mathcal{Y}; \widetilde{\theta}_{I_{k}, \tau}\right) - \ell_{I_{k}}\left(\mathcal{Y}; \widetilde{\theta}_{I_{0}, \tau}\right)\right|^{r} \IF(T_{1, \tau} > \mathfrak{z}_{1, \tau}) \leq \rho_{k}\mathcal{R}_{r}\left(\theta^{\ast}_{\tau}\right)
\end{equation}

Similarly, for every $q \geq 2$, the event $\mathcal{B}_{q}$ means that the first false alarm occurs at the step $q$ and $\widehat{\theta}_{\tau} = \widetilde{\theta}_{I_{q-1}, \tau}$. If $\mathfrak{z}_{1, \tau}, \cdots, \mathfrak{z}_{q-1, \tau}$ have already been fixed, the event $\mathcal{B}_{q}$ is only controlled by $\mathfrak{z}_{q, \tau}$, which is the minimal value that ensures
\begin{equation}
\label{critical2}
\displaystyle \max_{k \geq q} \operatorname{E}_{\theta^{\ast}_{\tau}}\left|\ell_{I_{k}}\left(\mathcal{Y}; \widetilde{\theta}_{I_{k}, \tau}\right) - \ell_{I_{k}}\left(\mathcal{Y}; \widetilde{\theta}_{I_{q-1}, \tau}\right)\right|^{r} \IF(\mathcal{B}_{q}) \leq \rho_{k}\mathcal{R}_{r}\left(\theta^{\ast}_{\tau}\right)
\end{equation}
Hence the value of $\mathfrak{z}_{q, \tau}$ can be obtained numerically by the Monte Carlo simulations for the nine different scenarios of fixed $\theta^{\ast}_{\tau}$. It is easy to prove that such defined $\mathfrak{z}_{q, \tau}$  fulfill the propagation condition \eqref{CARE_propagation_condition} in view of the decomposition \eqref{propoDecom}.
We summarize the concrete steps of calculating critical values,
\vspace{-0.30cm}
\begin{enumerate}
	\item select the minimum value satisfying \eqref{critical1} as the critical value of interval $I_{1}$,   $\mathfrak{z}_{1, \tau}$.
	\item Given $\mathfrak{z}_{1, \tau}$, select the minimum value satisfying \eqref{critical2} for $q=2$ as the critical value of interval $I_{2}$, $\mathfrak{z}_{2, \tau}$.
    \item Repeat step 2 for $q=3, \cdots, K$. Then we sequentially have $\mathfrak{z}_{k, \tau}$.
\end{enumerate}

%Steps for a fixed parameter, $\tau$, given simulated data series, risk bound
%\begin{enumerate}
%	\item Intervals selected
%	\item Obtain the test statistics of the simulated series
%	\item Sequential choice, Propagation condition
%	\item Find the largest(?) value...
%\end{enumerate}
%\vspace{0.50cm}

\begin{figure}[htp]
\begin{center}
\includegraphics[scale = 0.50]{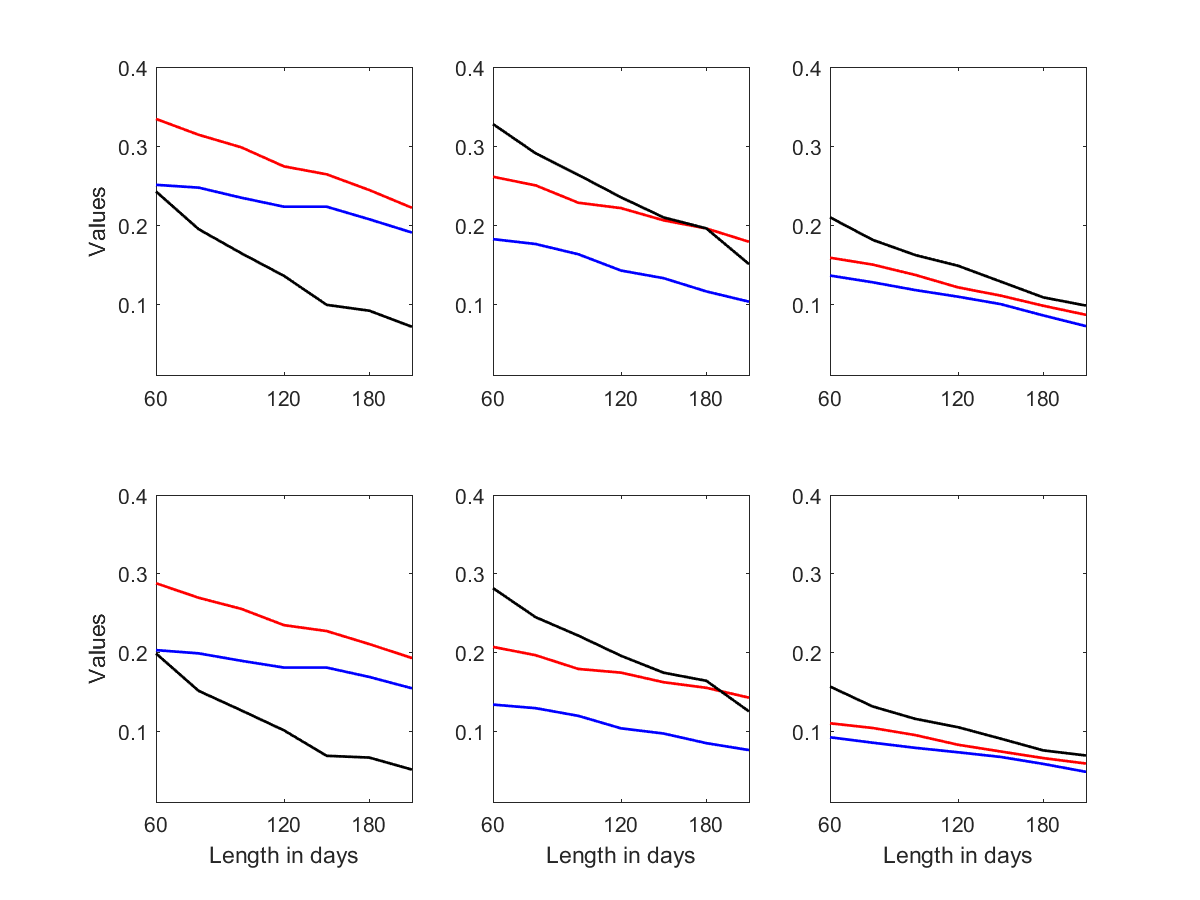}
\caption{Simulated critical values across different parameter constellations given in Table \ref{CARE_parameter_dynamics_quartiles} for the modest (upper panel, $r = 0.8$) and conservative (lower panel, $r = 1$) risk cases. We consider three expectile levels, $\tau = 0.05$ (blue), $\tau = 0.01$ (red) and $\tau = 0.0025$ (black).}
\label{CARE_6_critical_value_figure}
\end{center}
\end{figure}

The resulting critical value curves for the selected six 'true' parameter constellations from Table \ref{CARE_parameter_dynamics_quartiles} and associated risk bounds from Table \ref{CARE_risk_bound} are displayed in Figure \ref{CARE_6_critical_value_figure}. The upper (lower) panel represents critical values in the modest (conservative) risk case. The blue, red and black lines represents the expectile levels $\tau = 0.05$, $\tau = 0.01$ and $\tau = 0.0025$, respectively.

\textbf{Adaptive Estimation}

\begin{figure}[htp]
\begin{center}
\includegraphics[scale = 0.50]{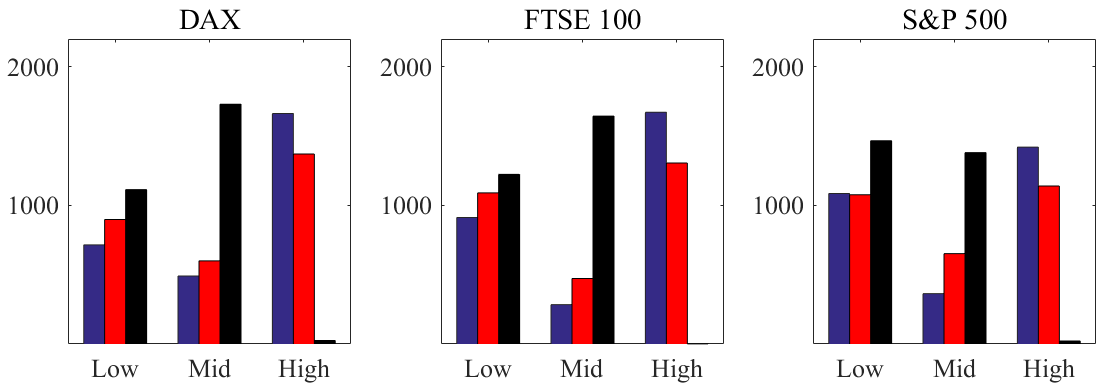}
\caption{Histogram of the selected parameter scenarios (Low, Mid and High) for adaptive estimation with $\tau = 0.05$ (blue), $\tau = 0.01$ (red), and $\tau = 0.0025$ (black).}
\label{CARE_Critical_Flag}
\end{center}
\end{figure}

Figure \ref{CARE_6_critical_value_figure} presents that critical values evolve in a decreasing route with a similar magnitude across all cases. When practicing the adaptive estimation, it is reasonable to choose the critical value set in a data-driven fashion: at a fixed time point, the yearly estimate $\widehat{\alpha}_{1, \tau}$ serves as a benchmark to select the appropriate scenario. If its value is, for example, lower (higher) than the reported first (third) quartile case in Table \ref{CARE_parameter_dynamics_quartiles}, then the corresponding left (right) panel of critical value curve is selected. Figure \ref{CARE_Critical_Flag} presents the frequencies of each critical value scenario for the three expectile level frequencies according to the closeness of parameter $\widehat{\alpha}_{1, \tau}$.

\textbf{Discussion}

In addition, one possible solution for obtaining critical values is to use the technique of multiplier bootstrap. Under this case, one can avoid simulating data based on an AND distribution assumption, which may be misspecified in practice. Some theoretical literatures have proved the validation under certain potential model misspecification with finite data sample, such as \citet{Spokoiny2015}, \citet{Spokoiny2015b}. The crucial idea is that the likelihood ratio test statistics between the parameter estimator and the true unknown parameter in the real world can be acceptably mimicked by the counterparts in the bootstrap world through multiplying the likelihood with a weight which is independent from the observations and generated using a distribution with mean and variance as 1. This method is totally data-driven. Under the assumption of independent observations, it can be theoretically proved that the confidence set in the bootstrap world can successfully represent that in real world even under a modest model misspecification. Unfortunately, there has not yet theoretical analysis in our autoregressive modelling situation. We refer to the extension of this technique in future researches.

On the other side, in order to test the validity of critical values using the simulation procedure, we further analyze its performance. When we allow the false alarm rate or significance level $\rho$,
$\PP(T_{k, \tau} > \mathfrak{z}^{\text{true}}_{k, \tau}) = \rho$, with $\mathfrak{z}^{\text{true}}_{k, \tau}$ denoted as the unknown true critical values for interval index $k$ and expectile level $\tau$. We practically use the simulated critical value $\mathfrak{z}_{k, \tau}$ as a substitute of $\mathfrak{z}^{\text{true}}_{k, \tau}$. Thus we can check the quality of approximation by investigating the difference $\delta = |\rho - \PP(T_{k, \tau} > \mathfrak{z}_{k, \tau})|$ with $\rho = 0.25$ as in the following figure \ref{CARE_Critical_Check}. Most of the differences $\delta$ are relatively small, largely lower than 5\%, and tend to decline as the interval length rises.
\begin{figure}[htp]
\begin{center}
\includegraphics[scale = 0.50]{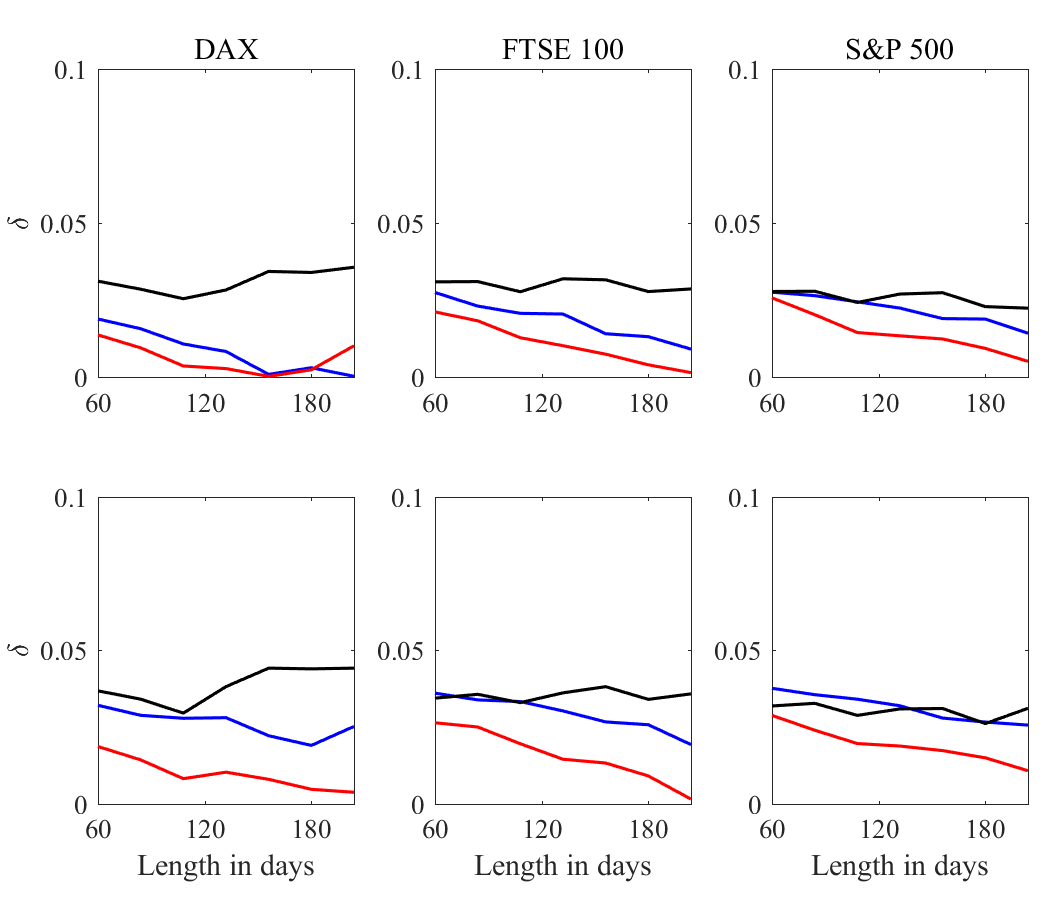}
\caption{Validation for the critical values for expectile level $\tau = 0.05$ (blue), $\tau = 0.01$ (red), and $\tau = 0.0025$ (black) with the modest (upper panel, $r = 0.8$) and conservative (lower panel, $r = 1$) risk cases. }
\label{CARE_Critical_Check}
\end{center}
\end{figure}

\subsection{Application}
\textbf{Multipliers of alternatives}
\vspace{-0.30cm}
\begin{figure}[htp]
\begin{center}
\includegraphics[scale = 0.5]{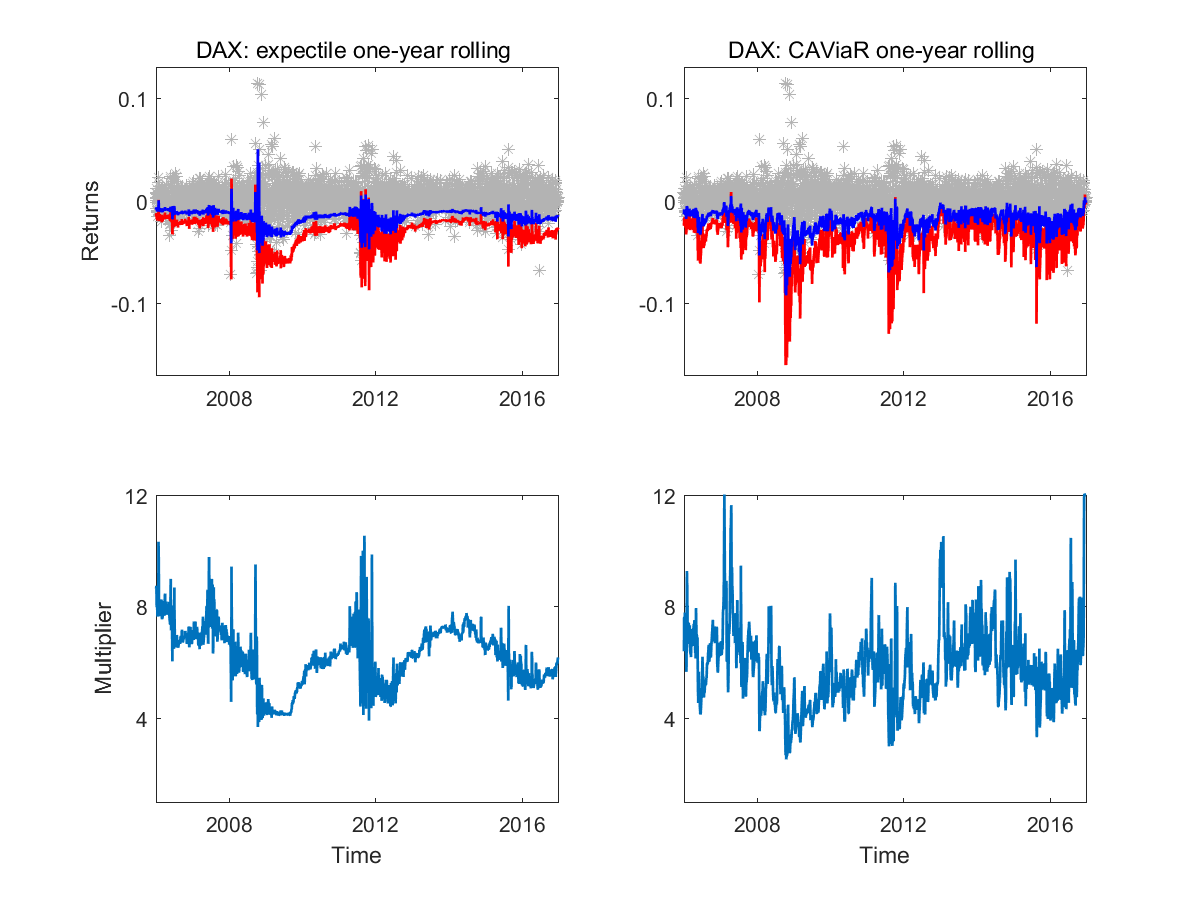}
\caption{Estimated expectile (blue) and expected shortfall (red) by one-year fixed rolling window (upper left panel), and the corresponding time-varying multiplier (lower left panel) for DAX index returns from 2 January 2006 to 30 December 2016. Also depicted are the estimated VaR (blue) ($\alpha = 0.065$) and expected shortfall (red) by CAViaR - based one-year rolling method (upper right panel), and the corresponding multiplier dynamics (lower right panel).}
\label{CARE_DAX_TIPP_rolling}
\end{center}
\end{figure}

\vspace{-0.30cm}
\begin{figure}[htp]
\begin{center}
\includegraphics[scale = 0.5]{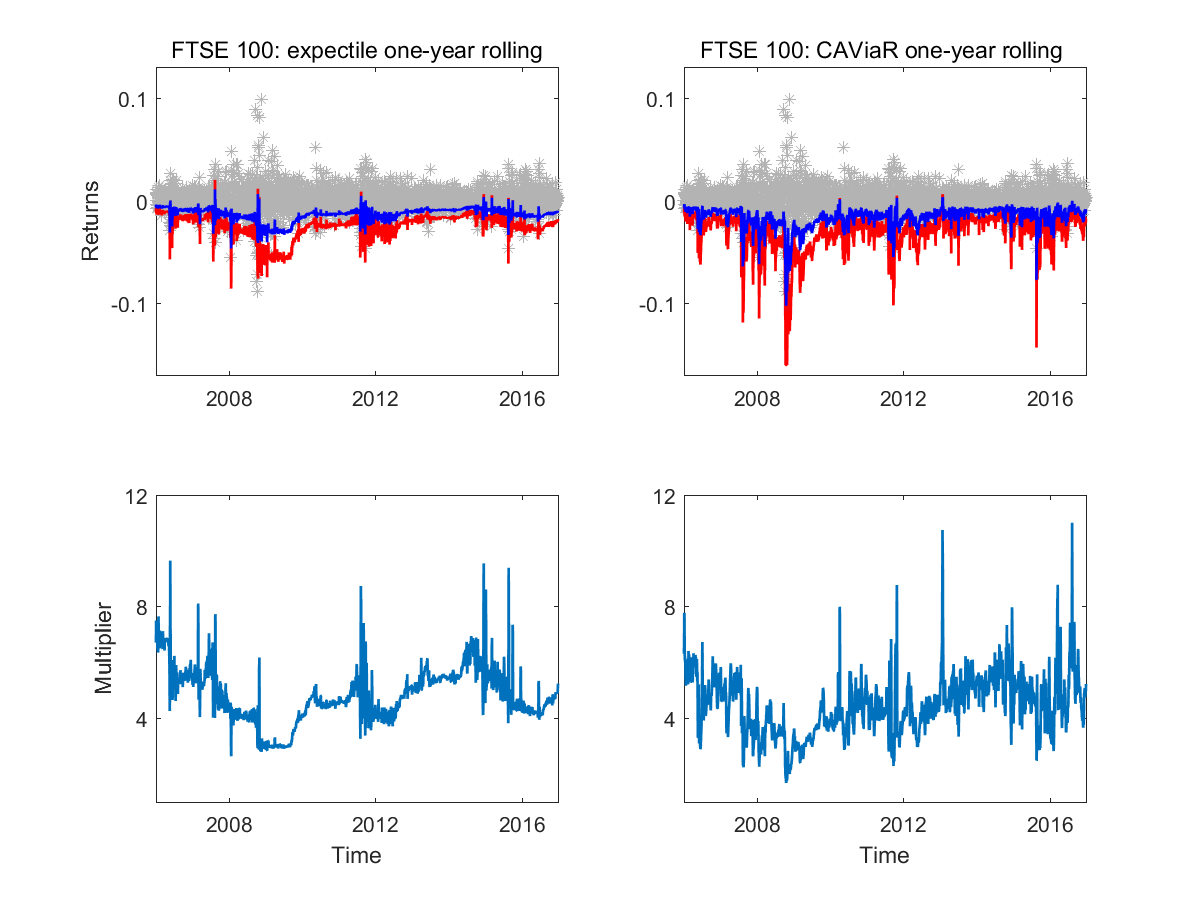}
\caption{Estimated expectile (blue) and expected shortfall (red) by one-year fixed rolling window (upper left panel), and the corresponding time-varying multiplier (lower left panel) for FTSE 100 index returns from 2 January 2006 to 30 December 2016. Also depicted are the estimated VaR (blue) ($\alpha = 0.065$) and expected shortfall (red) by CAViaR - based one-year rolling method (upper right panel), and the corresponding multiplier dynamics (lower right panel).}
\label{CARE_FTSE_TIPP_rolling}
\end{center}
\end{figure}

\vspace{-0.30cm}
\begin{figure}[htp]
\begin{center}
\includegraphics[scale = 0.5]{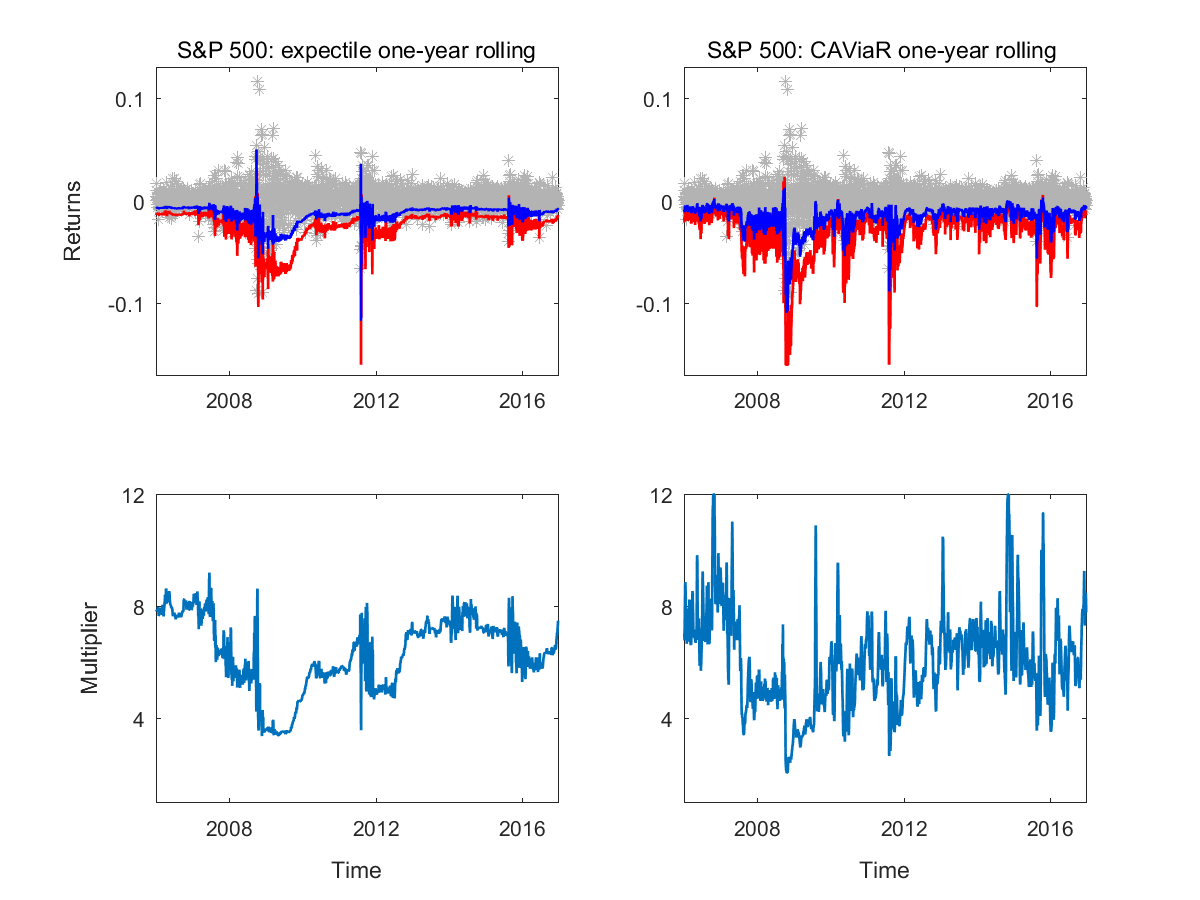}
\caption{Estimated expectile (blue) and expected shortfall (red) by one-year fixed rolling window (upper left panel), and the corresponding time-varying multiplier (lower left panel) for S\&P 500 index returns from 2 January 2006 to 30 December 2016. Also depicted are the estimated VaR (blue) ($\alpha = 0.065$) and expected shortfall (red) by CAViaR - based one-year rolling method (upper right panel), and the corresponding multiplier dynamics (lower right panel).}
\label{CARE_SP_TIPP_rolling}
\end{center}
\end{figure}

\textbf{Performance comparison}

\begin{table}[hp]
\begin{center}
\begin{tabular}{c|r@{.}lr@{.}lr@{.}lr@{.}lr@{.}lr@{.}l}
\hline\hline
\multicolumn{13}{c}{Panel A     DAX} \\
\hline
 & \multicolumn{2}{c}{Return(\%)} & \multicolumn{2}{c}{Volatility(\%)} & \multicolumn{2}{c}{VaR 99\%} &
 \multicolumn{2}{c}{Skewness} & \multicolumn{2}{c}{Kurtosis} & \multicolumn{2}{c}{Sharpe} \\
\hline
Multiplier 1   &     3&53  &    2&24  &   -0&40  &    0&11  &    9&15  &    0&10   \\
Multiplier 2   &     4&01  &    4&47  &   -0&81  &    0&09  &    9&12  &    0&06   \\
Multiplier 3   &     4&47  &    6&71  &   -1&22  &    0&08  &    9&04  &    0&04   \\
Multiplier 4   &     4&86  &    8&95  &   -1&65  &    0&06  &    9&02  &    0&03   \\
Multiplier 6   &     5&25  &   11&19  &   -2&07  &    0&05  &    9&01  &    0&03   \\
Multiplier 7   &     5&29  &   15&47  &   -2&79  &   -0&01  &    9&01  &    0&02   \\
Multiplier 8   &     2&02  &   10&39  &   -2&03  &   -0&06  &   42&25  &    0&01   \\
Multiplier 9   &     0&09  &   10&67  &   -2&21  &   -0&72  &   45&34  &    0&00   \\
Multiplier 10  &     0&01  &   10&82  &   -2&36  &   -1&56  &   39&92  &    0&00   \\
Multiplier 11  &     0&00  &   11&35  &   -2&51  &   -2&08  &   38&25  &    0&00   \\
Multiplier 12  &     0&00  &   11&95  &   -2&71  &   -2&04  &   35&83  &    0&00   \\ 
\hline

\multicolumn{13}{c}{Panel B     FTSE 100} \\
\hline
 & \multicolumn{2}{c}{Return(\%)} & \multicolumn{2}{c}{Volatility(\%)} & \multicolumn{2}{c}{VaR 99\%} &
 \multicolumn{2}{c}{Skewness} & \multicolumn{2}{c}{Kurtosis} & \multicolumn{2}{c}{Sharpe} \\
\hline
Multiplier 1   &     3&06  &    1&92  &   -0&31  &    0&02  &   10&65  &    0&10   \\
Multiplier 3   &     3&08  &    5&75  &   -0&97  &   -0&02  &   10&55  &    0&03   \\
Multiplier 4   &     3&04  &    7&67  &   -1&30  &   -0&03  &   10&61  &    0&03   \\
Multiplier 5   &     2&89  &    9&58  &   -1&63  &   -0&06  &   10&63  &    0&02   \\
Multiplier 6   &     1&07  &    8&39  &   -1&61  &   -0&78  &   22&21  &    0&01   \\
Multiplier 7   &     0&01  &    7&60  &   -1&72  &   -1&54  &   42&16  &    0&00   \\
Multiplier 8   &     0&00  &    8&19  &   -1&91  &   -1&02  &   39&20  &    0&00   \\
Multiplier 9   &     0&00  &    9&00  &   -2&11  &   -0&77  &   39&28  &    0&00   \\
Multiplier 10  &     0&00  &    9&86  &   -2&30  &   -0&66  &   38&90  &    0&00   \\
Multiplier 11  &     0&00  &   10&64  &   -2&55  &   -0&65  &   33&63  &    0&00   \\
Multiplier 12  &     0&00  &   11&11  &   -2&74  &   -0&52  &   23&67  &    0&00   \\ 
\hline

\multicolumn{13}{c}{Panel C     S\&P 500} \\
\hline
 & \multicolumn{2}{c}{Return(\%)} & \multicolumn{2}{c}{Volatility(\%)} & \multicolumn{2}{c}{VaR 99\%} &
 \multicolumn{2}{c}{Skewness} & \multicolumn{2}{c}{Kurtosis} & \multicolumn{2}{c}{Sharpe} \\
\hline
Multiplier 1   &     3&36  &    1&97  &   -0&37  &   -0&16  &   14&25  &    0&11   \\
Multiplier 2   &     3&69  &    3&95  &   -0&76  &   -0&17  &   14&26  &    0&06   \\
Multiplier 3   &     4&00  &    5&93  &   -1&12  &   -0&18  &   14&24  &    0&04   \\
Multiplier 5   &     3&90  &    9&56  &   -1&78  &   -0&38  &   14&80  &    0&03   \\
Multiplier 6   &     1&43  &    7&83  &   -1&70  &   -0&80  &   39&68  &    0&01   \\
Multiplier 7   &     0&08  &    7&28  &   -1&66  &   -2&66  &   53&56  &    0&00   \\
Multiplier 8   &     0&03  &    7&73  &   -1&86  &   -2&55  &   45&04  &    0&00   \\
Multiplier 9   &     0&00  &    8&42  &   -2&12  &   -2&24  &   42&20  &    0&00   \\
Multiplier 10  &     0&00  &    9&22  &   -2&33  &   -2&17  &   44&03  &    0&00   \\
Multiplier 11  &     0&00  &   10&05  &   -2&53  &   -2&13  &   45&73  &    0&00   \\
Multiplier 12  &     0&00  &   10&86  &   -2&77  &   -2&10  &   44&95  &    0&00   \\ 
\hline
\hline
\end{tabular}
\caption{Descriptive statistics of the portfolio returns based on the TIPP strategy under constant multipliers for the DAX index, FTSE 100 and S\&P 500 from 2 January 2006 to 30 December 2016. The investment strategy is based on a one-year investment horizon.}
\label{CARE_TIPP}
\end{center}
\end{table}

\end{document}